\pdfoutput = 1
\documentclass[apj,numberedappendix]{emulateapj}

\usepackage{epsfig}
\usepackage{amsmath}
\usepackage{enumerate}
\usepackage{mathrsfs}

\usepackage{natbib}
\usepackage{hyperref}

\usepackage{ulem}
\usepackage[squaren, Gray, cdot]{SIunits}

\usepackage{color}

\newbox\grsign \setbox\grsign=\hbox{$>$} \newdimen\grdimen \grdimen=\ht\grsign
\newbox\simlessbox \newbox\simgreatbox \newbox\simpropbox
\setbox\simgreatbox=\hbox{\raise.5ex\hbox{$>$}\llap
     {\lower.5ex\hbox{$\sim$}}}\ht1=\grdimen\dp1=0pt
\setbox\simlessbox=\hbox{\raise.5ex\hbox{$<$}\llap
     {\lower.5ex\hbox{$\sim$}}}\ht2=\grdimen\dp2=0pt
\setbox\simpropbox=\hbox{\raise.5ex\hbox{$\propto$}\llap
     {\lower.5ex\hbox{$\sim$}}}\ht2=\grdimen\dp2=0pt

\def\beq{\begin{equation}}
\def\eeq{\end{equation}}

\begin{document}

\title{The Thermodynamics of Rotating Black-Hole Star Clusters.}

\author{Andrei Gruzinov$^1$, Yuri Levin$^{2,3,4}$, and Jiarong Zhu$^1$}

\affil{$^1$CCPP, Department of Physics, New York University,
New York, NY 10001}
\affil{$^2$Physics Department and Columbia Astrophysics Laboratory, Columbia University, 
538 West 120th Street, New York, NY 10027}
\affil{$^3$Center for Computational Astrophysics, Flatiron Institute, 162 5th Ave, NY10011}
\affil{$^4$School of Physics and Astronomy, Monash University, Clayton VIC 3800, Australia}

\begin{abstract}
Rotating star clusters near supermassive black holes  are studied using Touma--Tremaine thermodynamics of gravitationally interacting orbital ellipses. A simple numerical procedure for calculating thermodynamic equilibrium states for an arbitrary distribution of stars over masses and semimajor axes is described.  Spontaneous symmetry breaking and breakdown of thermodynamics at low positive temperatures are rigorously proven for non-rotating clusters. Rotation is introduced through a second temperature-like parameter. Both axially symmetric and lopsided rotational equilibria are found; the lopsided equilibria precess with the angular velocity that is given by the ratio of the two temperatures. Eccentric stellar disc in the nucleus of Andromeda galaxy may be an example of a lopsided thermodynamic equilibrium of a rotating black hole star cluster. Stellar-mass black holes occupy highly eccentric orbits in  broken-symmetry star clusters, and form flattened disc-like configurations in rotating star clusters. They are attracted to orbits that are stationary in the frame of reference rotating with the angular velocity of the cluster. In spherical clusters, stellar-mass black holes' orbits are significantly more eccentric than those of the lighter stars if the temperature is negative, and more circular if the temperature is positive. Finally we note that planets, comets, dark matter particles and other light bodies tend to form a spherically symmetric non-rotating sub-cluster with maximum-entropy eccentricity distribution $\mathscr{P}(e)=2e$, even if their host cluster is rotating and lopsided.

\medskip
\end{abstract}


\section{Introduction}

An old subject of orbital dynamics in near-Keplerian potentials has been revived over the past two decades, in order to achieve a deeper understanding of dynamics of stellar-mass objects near supermassive black holes in galactic nuclei. Since the work of \cite{1996NewA....1..149R}, it has been understood that secular, orbit-averaged interactions between the stellar orbits play dominant role in determining the evolution of angular momenta and eccentricities of the orbits. The relatively fast secular dynamics leaves  semimajor axes of the orbits unchanged; the axes evolve on a much longer timescale due to 2-body gravitational scattering of the stars.  It is thus of considerable interest to explore a purely secular evolution of black-hole star clusters. 

The original insight has since been complemented by a large set of numerical and analytical exploration of the secular dynamics \citep{2007MNRAS.379.1083G,2011ApJ...738...99M,2011PhRvD..84d4024M, 2014MNRAS.443..355H, 2016ApJ...820..129B, 2016MNRAS.458.4129S, 2016MNRAS.458.4143S, 2018ApJ...860L..23B, 2018MNRAS.481.4566F}. The purpose of these works was to find an effective description of the stochastic evolution of orbital parameters of individual stars, dubbed ``resonant relaxation'' by \cite{1996NewA....1..149R}. 

The outcome of resonant relaxation was studied 
 in a series of papers  \cite{2014JPhA...47C2001T}, \cite{2019PhRvL.123b1103T}, \cite{2019MNRAS.tmp.2761T}, and \cite{2020MNRAS.493.2632T} (TT). TT argued that secular dynamics allows equilibria states that can be described by a language of conventional statistical mechanics, with temperature $T$ serving as a measure of self-gravitation energy of the cluster. It is convenient to define $\beta=1/T$; ${\rm TT}$ show that $\beta$ can be both positive and negative. Remarkably, while non-rotating low-$\beta$  equilibria are spherically symmetric, the high-$\beta$ (low positive temperature) equilibria turned out to be non-spherical. This phase transition and the associated lopsided gravitational potentials and stellar configurations have important practical implications for the stellar and gas dynamics near supermassive black holes.
 However, ${\rm TT}$ demonstrated this behavior only for special cases (one article per each case) and did not give general proof for the existence of the phase transition.
 
 This paper advances ${\rm TT}$'s  discovery in several ways. Firstly, we devise a simple 
 numerical algorithm that allows us to compute thermal equilibria of stellar clusters with any distribution of semimajor axes and stellar masses. 
 
 Secondly, we give a general proof for the existence of the phase transition and elucidate the limits of applicability for thermodynamical treatment, pointing out that the full partition function diverges for sufficiently high $\beta$ (low positive temperatures). 
 
 Thirdly, we add rotation to the cluster [this was done in Touma \& Tremaine (2014) for discs but not for 3-dimensional clusters], and numerically explore both low-$\beta$ axially symmetric equilibria, and high-$\beta$ lopsided equilibria that precess with a fixed angular velocity. We demonstrate the latter configurations that appear similar to the eccentric nuclear stellar disc in Andromeda \citep{1995AJ....110..628T}. 
 
 Fourthly, we explore the equilibrium configurations of stellar-mass black holes that are much heavier than average members of the cluster. We find, e.g.,  that they cluster on strongly eccentric orbits in lopsided non-rotating equilibria and that they form a strongly flattened disc-like structures in rotating clusters.
 
 The plan of the paper is as follows. In Section 2 we  describe the general formalism for thermodynamic equilibria of black hole clusters. In Section 3 we give a proof for spontaneous symmetry breaking in non-rotating clusters and describe the limits of applicability of thermodynamics. In Section 4 we describe the numerical algorithm for constructing equilibria and in Section 5 we present results of our numerical experiments. In Section 6 we explore analytically the distributions of stellar mass black holes in both spherical and rotating clusters. We also comment on the universality of distribution of light objects. We conclude in Section 7 by briefly discussing possible astrophysical implications of our findings.

\section{Nonlinear Poisson Equation}\label{seq:pois}

The secular-dynamical equilibrium  state of a black-hole star cluster is achieved by evolution of the stellar Keplerian ellipses, in which the semimajor axes remain unchanged, while all other orbital parameters relax, preserving only the integrals of motion. Therefore at least one thermodynamic equilibrium state must exist for any set of quantities $(F,U,{\bf J})$, where
\begin{itemize}

\item $F(A)$ is the distribution function of stars over masses $m$ and semimajor axes $a$.  We define a composite $A\equiv(m,a)$, with the number of stars $N=\int dA~F(A)$,  $dA\equiv dm~da$. 

\item $U$ is the potential energy of gravitationally attracting ellipses. The mass of each star is spread over its ellipse in proportion to the orbital time, as spelled out below. 

\item ${\bf J}$ is the total angular momentum of the stars.  

\end{itemize}

For a given set $(F,U,{\bf J})$, in the mean-field approximation, the thermodynamic equilibrium state is characterized by 
\begin{itemize}

\item $\phi({\bf r})$  -- the equilibrium gravitational potential of the stars only (the Keplerian potential of the black hole not included) 

\item $f(A,B)$ -- the equilibrium distribution function of stars over masses $m$ and semimajor axes $a$, eccentricities $e$,  and ellipse orientations, given by unit vectors along the major and the minor axes $\hat{n}_1$, $\hat{n}_2$. Here we have introduced another composite variable $B\equiv(e,\hat{n}_1,\hat{n}_2)$
The total number of stars 
is given by
\begin{equation}
    N=\int dA~dB~f(A,B),
\end{equation}
where
\beq
dB\equiv de^2~d^2n_1~d^2n_2~\delta(\hat{n}_1\cdot \hat{n}_2)
\eeq
and $d^2n_{1,2}$ are the differential solid angles. 
\end{itemize}

In statistical physics language, the cluster can be represented by a
micro-canonical ensemble with two additive conserved quantities,
energy and angular momentum. Therefore, in the mean field theory
approximation,  the canonical equilibrium distribution function has
Boltzmann-like factors for both energy and angular momentum [see also Touma \& Tremaine (2014) for derivation using the
maximum-entropy argument]. It is
given by
\beq\label{eqdi}
f(A,B)={F(A)\over Z(A)}\exp\left[{-\beta u(A,B)}+\vec{\gamma}\cdot {\bf j}(A,B)\right], 
\eeq
where 
\begin{itemize}

\item $u(A,B)$, {\bf j}(A,B) are the gravitational potential energy due to the gravitational field from other ellipses, and the angular momentum of the $(A,B)$ ellipse. They are given by
\beq
{\bf j}(A,B)=\left[GMm^2a(1-e^2)\right] ^{1/2}\hat{n}_1\times \hat{n}_2,
\eeq
where $M$ is the black hole mass, and
\begin{eqnarray}
u(A,B)=m\langle\phi\rangle&=&\frac{m}{P}\int\limits_{0}^{2\pi}d\xi~\frac{dt}{d\xi}~\phi({\bf R})\label{orbitenergy}\\
&=&{m\over 2\pi a}\int\limits_{0}^{2\pi} d\xi~R~\phi({\bf R})\nonumber
\end{eqnarray}

Here $\xi$ is the eccentric anomaly of a point on the Keplerian ellipse,  ${\bf R}(\xi)$ and $t(\xi)$ are the corresponding position and time from the periastron passage, and $P$ is the orbital period. These quantities are given by  
\begin{eqnarray}
P(A)&=&2\pi(GM)^{-1/2}a^{3/2},\\
t(\xi; A,B)&=&\frac{P(A)}{2\pi}(\xi -e\sin\xi ),\\
{\bf R}(\xi; A,B)&=&a(\cos\xi -e)\hat{n}_1+a\sqrt{1-e^2}\sin\xi \hat{n}_2.
\end{eqnarray}

\item As defined in the Introduction, $\beta$ is the inverse temperature. It can be either positive or negative, since the phase space of Keplerian ellipses with fixed semimajor axis is compact\footnote{The possibility of the temperature being negative for systems with compact phase spaces was first pointed out by \cite{1949NCim....6S.279O}.}.

\item $\vec{\gamma}$ is a 3-dimensional vector of inverse temperature-like
quantities corresponding to the components of angular momentum ${\bf J}$. 
For non-zero $\beta$, the factor in the exponential can be re-written as
$-\beta u_{J}$, where 
\begin{eqnarray}
    u_J&\equiv& u-\vec{\Omega}\cdot {\bf j}\label{jacoby}\\
    \vec{\Omega}&\equiv& \vec{\gamma}/\beta.\nonumber 
\end{eqnarray}
The quantity $u_J$ has the form of the Jacobi integral, a conserved quantity in a potential that is rotating with angular velocity $\vec{\Omega}$. By Jean's theorem, the steady-state distributions in such rotating frame should be a function only of
$u_J$. Therefore, if the solution we find is non-axisymmetric with respect to $\vec{\gamma}$, it should be interpreted as a 
solution that is obtained in a frame that is rotating with the angular velocity $\vec{\Omega}$ (we thank Scott Tremaine for clarifying this point). As we show below, a precessing
eccentric nuclear disc in Andromeda is a possible example of such solution. Conversely, for sufficiently ``hot'' systems with large $\vec{\Omega}$, no non-axisymmetric solutions can exist: it would be unphysical for a lopsided system to precess  with large angular velocity.

In actual numerical calculations we use
 $\cosh\left[\vec{\gamma}\cdot {\bf j}\right]$ rather than an $\exp\left[\vec{\gamma}\cdot {\bf j}\right]$ for the angular momentum Boltzmann factor because any ellipse can be traced in two opposite directions. This allows us to use 
 non-oriented ellipses and save on the configuration space sampling.

\item $Z(A)$ is the statistical sum, which must be calculated for each set of parameters $A$ separately, because $A={\rm const}$ during the secular-dynamical relaxation of the cluster:
\beq
Z(A)=\int dB~e^{-\beta u(A,B)}\cosh\left[\vec{\gamma}\cdot {\bf j}(A,B)\right],  
\eeq
\end{itemize}

The gravitational potential of the ellipses is given by the nonlinear Poisson equation 
\beq
\label{poiss}
\nabla ^2\phi=4\pi G\rho, 
\eeq
where the density $\rho$ is given by
\begin{eqnarray}
\rho({\bf r})&=&\int dA~dB~f(A,B)~\frac{m}{P(A)}\times\label{density}\\
& &\int\limits_{0}^{2\pi}d\xi~\frac{dt(\xi;A,B)}{d\xi}~\delta \left[ {\bf r}-{\bf R}(\xi;A,B) \right]\nonumber
\end{eqnarray}
The Poisson equation is nonlinear because the distribution function $f(A,B)$ nonlinearly depends on the gravitational potential $\phi$.

\section{Proof of Spontaneous Symmetry Breaking. The minimal temperature phenomenon. }\label{seq:sym}

There are two remarkable features of  the black-hole star cluster thermodynamics: 1. The spontaneous symmetry breaking at low temperatures, that has been demonstrated for particular configurations in ${\rm TT}$, and \newline 2.
The existence of positive ``minimum temperature'' $T_{\rm c}$, which is described here for the first time.  For values $\beta>1/T_{\rm c}$, the full statistical sum diverges and the function $f(A,B)$ collapses to a singular distribution. We note that such singular distributions have infinite binding energy. We emphasize that the divergence of the full statistical sum does not imply that no microcanonical ensemble with finite energy and mean-field 
Botzmann distributions  with $\beta>1/T_{\rm c}$ can exist. It does imply that if the cluster interacts with the heat bath with $\beta>1/T_{\rm c}$, it will collapse to a degenerate state.

The existence of the symmetry breaking and of the ``minimum temperature'' are described  analytically and rigorously proven in this section. The qualitative understanding of BH star clusters gives us confidence that our numerical results should be correct, as we do see both the symmetry breaking and the  low-temperature singularity in \S\ref{seq:numres}. Our analytical proof  is much simpler and more universal than the arguments in ${\rm TT}$. 

We are able to prove the spontaneous symmetry breaking in non-rotating clusters only, with $\gamma=0$ in Eq.~(\ref{eqdi}); in other words we are able to prove the breaking of spherical symmetry. Axial symmetry breaking occurs in rotating clusters too, since firstly, by continuity we expect it to take place at small $\gamma$ and secondly, we observe it in numerical simulations. Still, our proof works only for the non-rotating clusters. 

The minimal temperature phenomenon, i.e.~the breakdown of Touma--Tremaine thermodynamics at sufficiently low temperatures, is valid and proved below for clusters with arbitrary rotation. This proof is an immediate extension of Lemma (1) of the symmetry breaking proof. 

The spherical symmetry breaking follows from two observations: 
\newline
{\bf Lemma (1)}: For any given distribution $F(A)$, assumed "nice" enough, there exist initial (thermodynamically unrelaxed) distributions $f(A,B)$  with arbitrarily large binding energy $|U|$. 
\newline 
{\bf Lemma (2)}:  For any given distribution $F(A)$, all spherically symmetrical states have binding energy below a certain maximal value.  It follows that the only way to cool down the cluster, that is to increase the binding energy, is to break the spherical symmetry. We now prove (1) and (2) in turn. 

To prove Lemma (1), assume that all orbital ellipses are degenerate, with $e=1$, and aligned along single direction $x$. Then $U=-\infty$, because \begin{equation}
    \int \frac{dx_1~dx_2~\chi(x_1)\chi(x_2)}{|x_1-x_1|}=\infty.
\end{equation}
Here $\chi(x)$ is the linear density along $x$.

The divergence is logarithmic in $x$ and therefore also logarithmic in the eccentricity deviation from unity and in the misalignment angle of different ellipses. This leads to the interesting minimal-temperature phenomenon: Touma--Tremaine thermodynamics
breaks down at small positive temperatures,  because the full statistical sum $Z=\int dA~F(A)Z(A)$ diverges algebraically for $\beta>\beta_{\rm c}>0$. To prove the statement and to get an estimate of the critical temperature $T_c\equiv\beta_c^{-1}$, consider nearly degenerate ellipses, $e=1-\epsilon$, $\epsilon\ll 1$, which are nearly aligned, that is the ellipses have major axes directions within a cone of opening angle $\theta\ll 1$. For convenience we assume that the ellipses have similar semimajor axes $\sim a$ (this assumption is easy to relax but facilitates exposition of the main point). Since the minor axes of the ellipses are $\sim \epsilon^{1/2}a$, all the mass of the stars lies within a cylinder of length $\sim a$ and radius $\sim \max(\epsilon^{1/2},\theta)a$. Then the  self-gravitational energy of $N$ stars of mass $\sim m$ is, to logarithmic accuracy,
\beq
U\sim \frac{GN^2m^2}{a}\ln {\max(\epsilon^{1/2},\theta)}.
\eeq
The phase space volume of our nearly aligned and almost degenerate ellipses is $V_{\rm ph}\propto (\epsilon \theta^2)^N$. The contribution of these ellipses to the statistical sum (exact, not the mean-field) scales as
\begin{equation}
\propto V_{\rm ph}e^{-\beta U}
\end{equation}
and diverges for small $\epsilon$, $\theta$ if and only if  $\beta>\beta_{\rm c}$, 
\beq
\beta_{\rm c}^{-1}\equiv T_c\sim \frac{GNm^2}{a}.
\eeq
When positive temperature is lowered below $T_c$, the full distribution function should collapse to degenerate ellipses
\footnote{We emphasize again that the divergence of the full statistical sum and the existence
of the mean-field Poisson-Boltzmann states are not in a one-to-one
correspondence. It is possible 
that the broken symmetry mean-field thermodynamic equilibria, although
they do correctly describe the actual physical states of BH star
clusters, correspond to temperatures below $T_c$, when the full
statistical sum actually diverges.  We are working on clarifying this point.}. 
Note that $T_c$ corresponds to typical binding energy of a star to the cluster, and is thus comparable to a natural temperature scale of the cluster.  The collapse is readily observed in our numerical simulations, as described in \S\ref{seq:numres}.
We note that physically the collapse to a degenerate state can take place if the cluster interacts with the heat bath that is able to absorb a formally infinite amount of the degenerate state's binding energy. 

Lemma (2) is most easily proved by recalling that in a spherical black hole star cluster, an
elliptical orbit precesses  
in a direction that is retrograde with respect to its orbital motion\footnote{Not including relativistic precession, which is prograde.}.
This statement is proved in section 3.2 of \cite{2005ApJ...625..143T}. 
The angular frequency of the precession equals $\partial u/\partial j$, and ``retrograde'' implies that this is $<0$. Here $u$ is the orbit-averaged potential energy of the star, $j$ is the magnitude of its angular momentum, and the derivative is evaluated while keeping the orbital semi-major axis fixed.
It follows that the energy of the orbit is reduced as the orbit becomes more
circular. Applying this to all orbits at the same time, we see that for a
given $F(A)=F(a,m)$ the gravitational energy $U$ of the cluster is minimized (and its binding energy is maximized) if all orbits are circular.
This minimal energy is given by
\begin{equation}
    U_{\rm min}=-\int {da} {GM_{\rm cluster}(<a)\over a}\int dm~mF(a,m),
\end{equation}
where $M_{\rm cluster}(<a)$ is the stellar mass inside radius $a$:
\begin{equation}
    M_{\rm cluster}(<a)=\int_0^a da_1\int dm_1 m_1 F(a_1,m_1).
\end{equation}

As we saw from Lemma (1), there are cluster configurations with energies smaller than $U_{\rm min}$. 
They must have broken spherical symmetry. In our numerical experiments in \S\ref{seq:numres}, we demonstrate
 symmetry breaking at low positive temperature, and collapse to aligned degenerate  ellipses at an even smaller positive temperature. Before we show these results, we discuss our computational technique in the following section.

\section{The Numerical Method}
${\rm TT}$ solve the nonlinear Poisson equation (\ref{poiss}) using various simplifying    assumptions and series expansions. 
In this section we show that a direct brute force solution of the nonlinear system of equations (\ref{eqdi},\ref{poiss}) is possible, with  only minor numerical inventiveness. The numerical method is described below, the results -- in Section 5.

The computations shown in Section 5 require a few-minute calculation on a laptop to find an equilibrium state starting from an arbitrary distribution, and much less time to find a nearby equilibrium. A typical phase space covering  used in our computations was as follows. For parameters $A$ specifying the masses and semimajor axes, we typically use  $N_m=1$ (all stars have the same mass), and  $N_a=30$ possible values of the semimajor axis. For parameters $B$ specifying the orbital ellipses, we use $N_e=30$ values of eccentricity distributed uniformly in $e^2$ between $0$ and $1$, $N_1=600$ directions of major axes $\hat{n}_1$ on a Fibonacci spherical lattice, and $N_2=30$ perpendicular directions of minor axes $\hat{n}_2$, uniformly distributed over the angle of just $\pi$, rather than $2\pi$, since an ellipse is traced in both directions in the distribution function given by Eq.(\ref{eqdi}). We represent each ellipse by $N_\xi=50$ of its points that are uniformly distributed in the eccentric anomaly $\xi$ between $0$ and $2\pi$, and are weighted by a factor $\propto (1/P)m~dt/d\xi=(2\pi)^{-1}m~R/a$. This is done in order to compute the potential energy of the ellipse, as well as the mass density distribution created by all of the ellipses. The gravitational potential and the density are defined on $N^3$, $N=151$,  regular spacial grid, and each point representing each ellipse is assigned to a grid cell. 

The numerical procedure is as follows. Fix the inverse temperatures $\beta$ and $\vec{\gamma}$, and the distribution function $F(A)$.  For finding an equilibrium state for the first time,  start with an arbitrary initial potential $\phi$. For finding an equilibrium state that is close to the one previously found, but with slightly altered parameters, start with the previously calculated potential $\phi$. The computation proceeds iteratively, by repeating the following steps until the potential $\phi$ converges (i.e. does not change significantly  between successive iterations):

\begin{enumerate}

\item Given $\phi$: Fix $A$, calculate the weights $w=e^{-\beta u(A,B)}\cosh\left[\vec{\gamma}\cdot {\bf j}(A,B)\right]$ for all of the ellipse eccentricities and orientations $B$, simultaneously calculating the statistical sum $Z(A)=\sum w$. Repeat for all $A$ and obtain $f(A,B)$ from Eq.~(\ref{eqdi}); these are the weight factors for the ellipses.

\item Given $f$: Calculate $\rho({\bf r})$ on the spacial grid, by using Eq.~(\ref{density}) and replacing the integrals with sums. Calculate several lowest multipoles of $\rho$; we found it sufficient to compute the dipole, quadrupole, and octupole moments.

\item Given $\rho$: Calculate $\phi$ from the Poisson equation~(\ref{poiss}). We have used a simple relaxation method, by numerically solving the evolution equation $\partial _{\bar{t}}\phi=\nabla^2\phi-4\pi G\rho$, where $\bar{t}$ is the auxiliary time. The boundary conditions at the faces of the computation cube $N^3$ are given by the multipole expansion of $\phi$ using the multipoles of $\rho$ computed in the previous step. The size of the cube was chosen to be $4$ times greater than the size of the largest semi-major axis of a star in our sample. The number of $\bar{t}$-steps was chosen so as to make the $\rho$ updating  steps (1),(2) as computationally expensive as the $\phi$ updating step (3); typically 
$\sim 1000$ $\bar{t}$-steps per one density update. It takes $\sim N^2 \approx 20,000$ $\bar{t}$-steps for the potential relaxation procedure to converge. So, the procedure converges after a few dozen density updates.

\end{enumerate}

The numerical convergence was tested by (1) repeating the calculations  at
different resolutions, (2) by comparing the numerical results to  a
few analytically doable calculations, (3) by comparing the numerical
results to  high-resolution spherically symmetrical numerical results,
as explained in the next section.

\begin{figure}[h!]
\centering
\includegraphics[width=.51\textwidth]{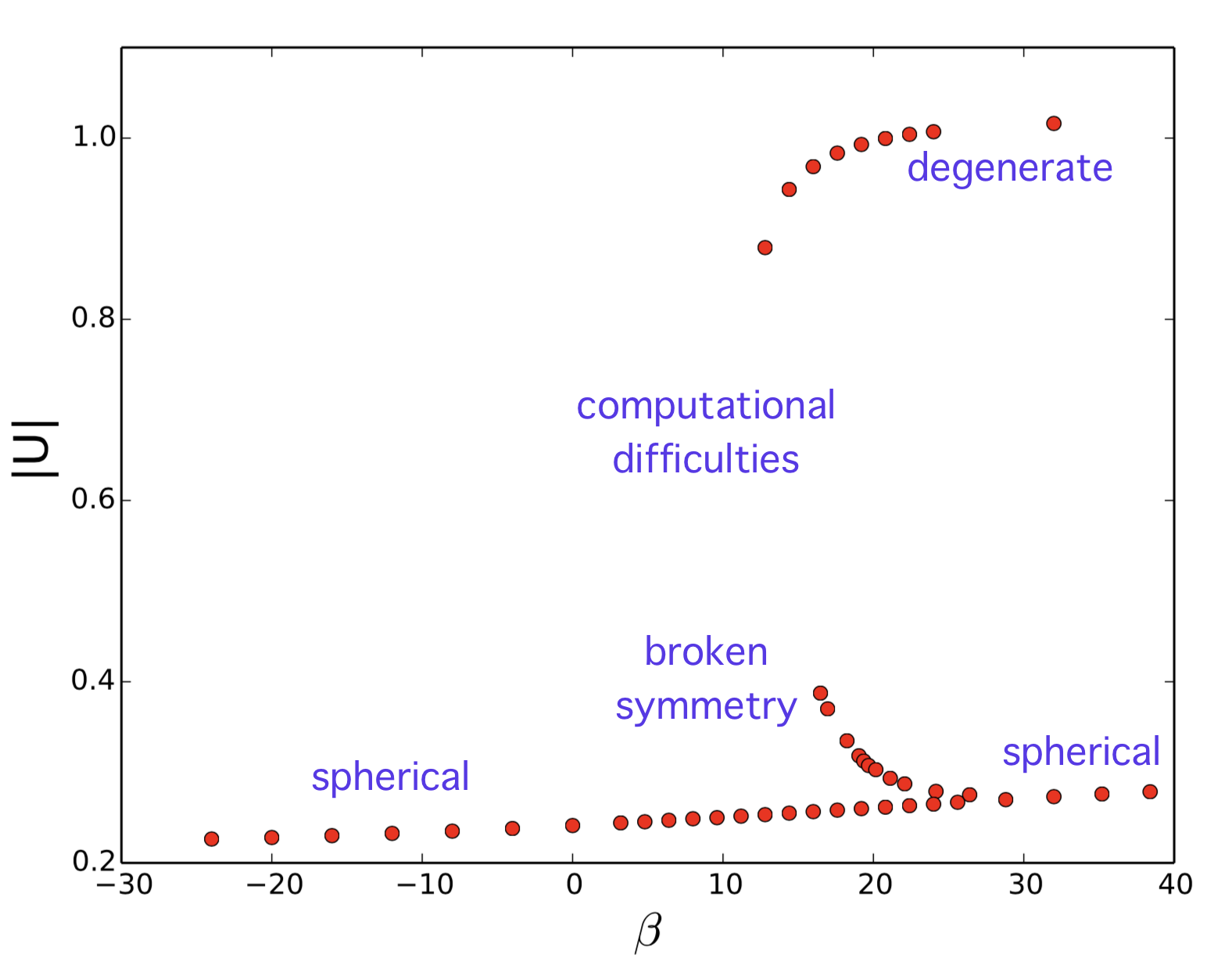}
\caption{\label{f1} Thermodynamic equilibria in  $\beta-|U|$ plane for a non-rotating cluster. The cluster is made of stars with the same mass $m_0$, with the total cluster mass $M_{\rm cluster}$ and semimajor axes uniformly distributed in the interval $(a_0,2a_0)$. $|U|$ is in units of $GM_{\rm cluster}^2/a_0$, $\beta$ is in units of $(GM_{\rm cluster}m_0/a_0)^{-1}$.}
\end{figure}

\begin{figure}[h!]
\centering
\includegraphics[width=.51\textwidth]{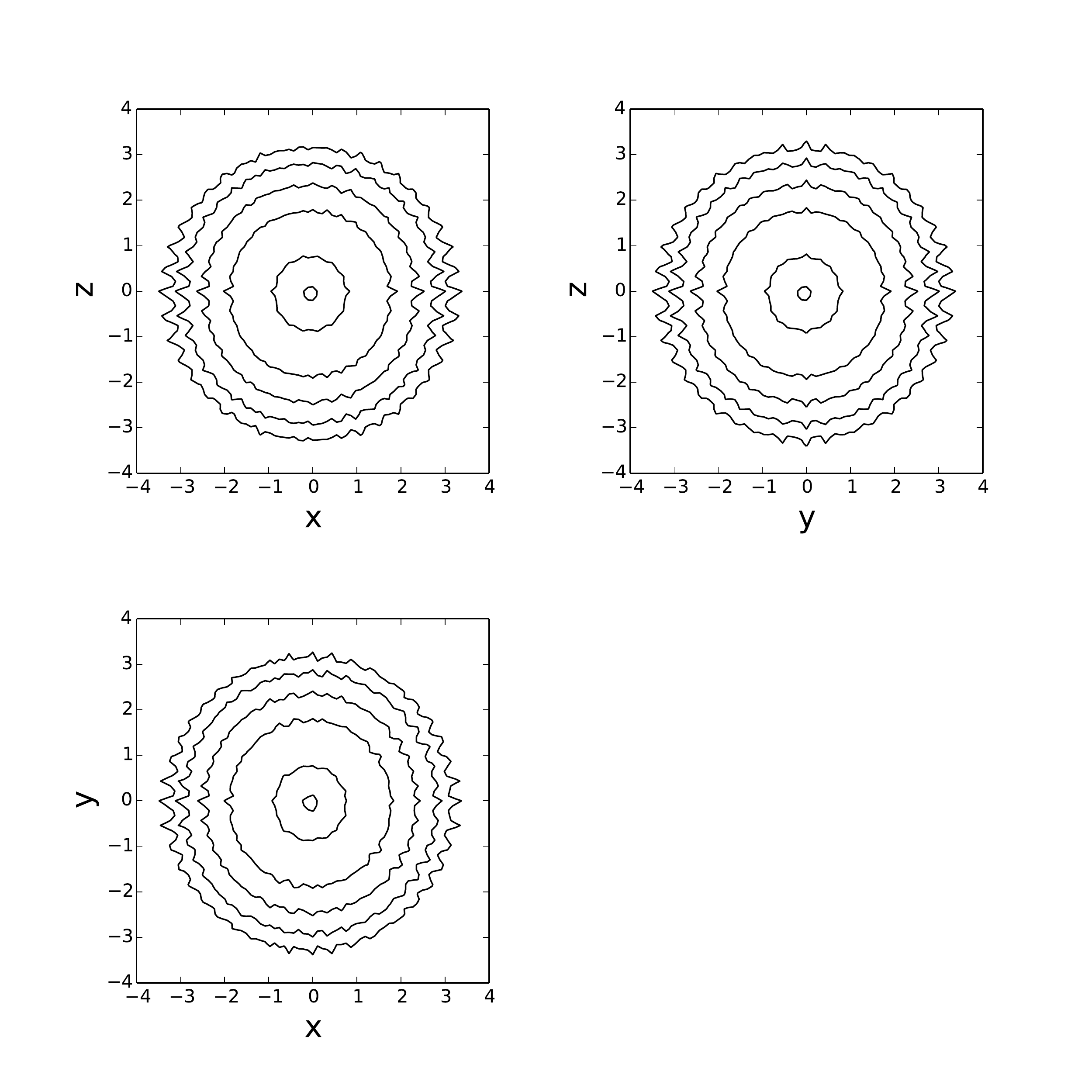}
\caption{\label{f24} Spherical low-$\beta$ ``hedgehog'' state. The stars are on eccentric orbits. Projected surface density  for $\beta=-24.0$, $\gamma=0$,  $|U|=0.226$. The isolines are 0.025, 0.05, 0.1, 0.2, 0.4, 0.8 of the maximal projected surface density. The ruggedness is due to the finite number of ellipse orientations $(600)$ used in the numerical procedure.}
\end{figure}

\begin{figure}[h!]
\centering
\includegraphics[width=.51\textwidth]{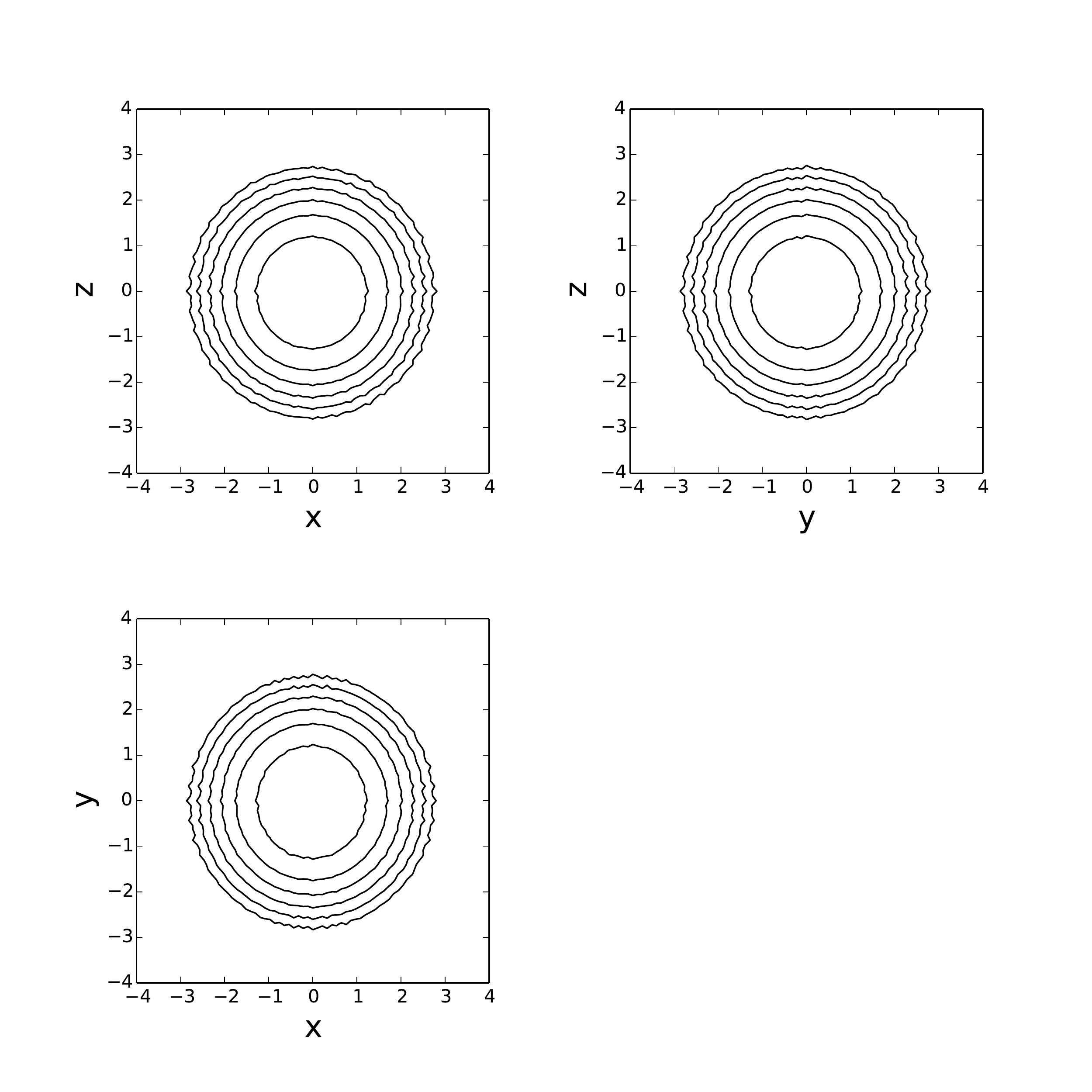}
\caption{\label{f35} Spherical high-$\beta$ circular-orbit state. Projected surface density  for $\beta=35.7$, $\gamma=0$,  $|U|=0.276$.}
\end{figure}

\begin{figure}[h!]
\centering
\includegraphics[width=.51\textwidth]{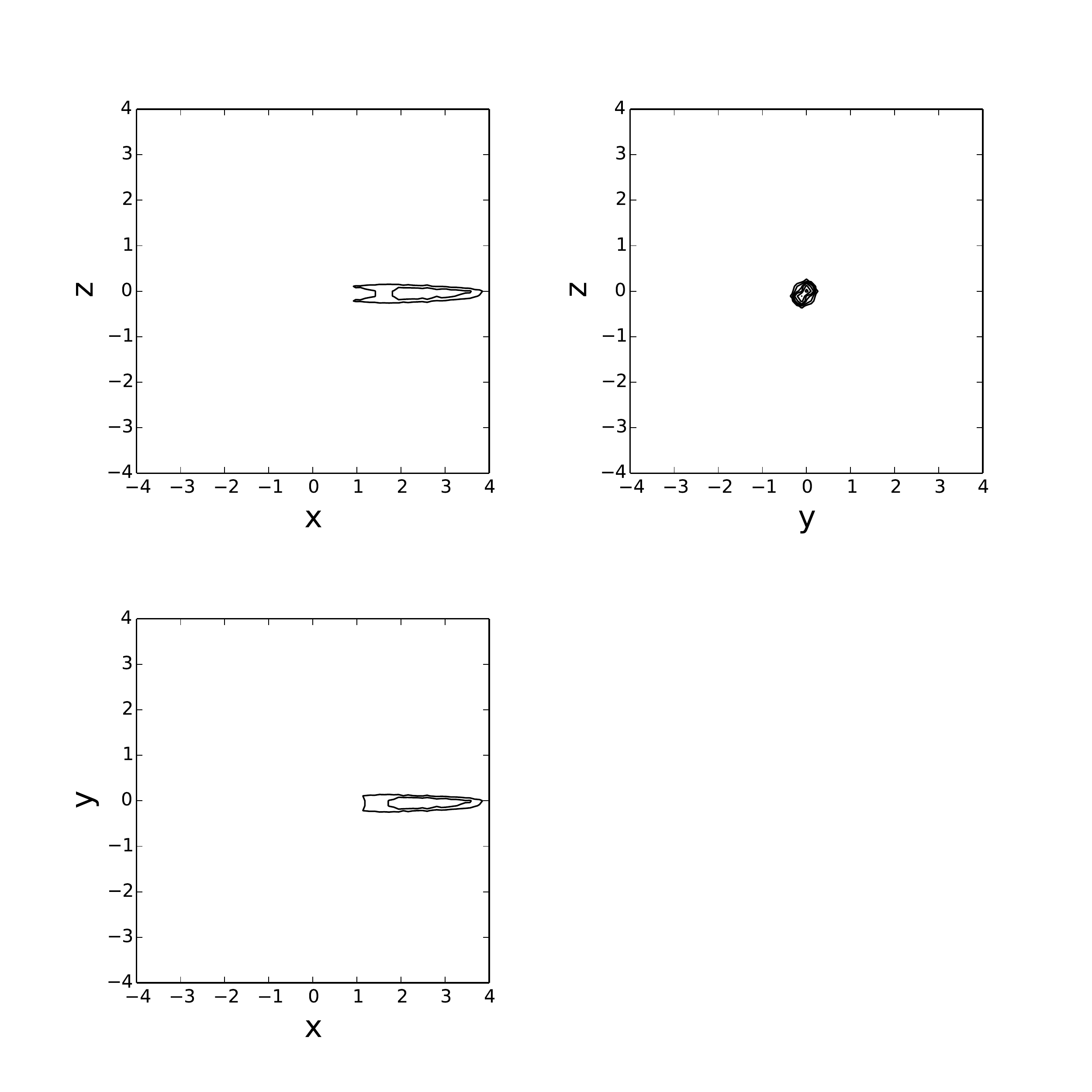}
\caption{\label{f13} Degenerate state. Projected surface density for $\beta=12.8$, $\gamma=0$, dipole moment $d=2.12$, $|U|=0.88$.}
\end{figure}

\begin{figure}[h!]
\centering
\includegraphics[width=.51\textwidth]{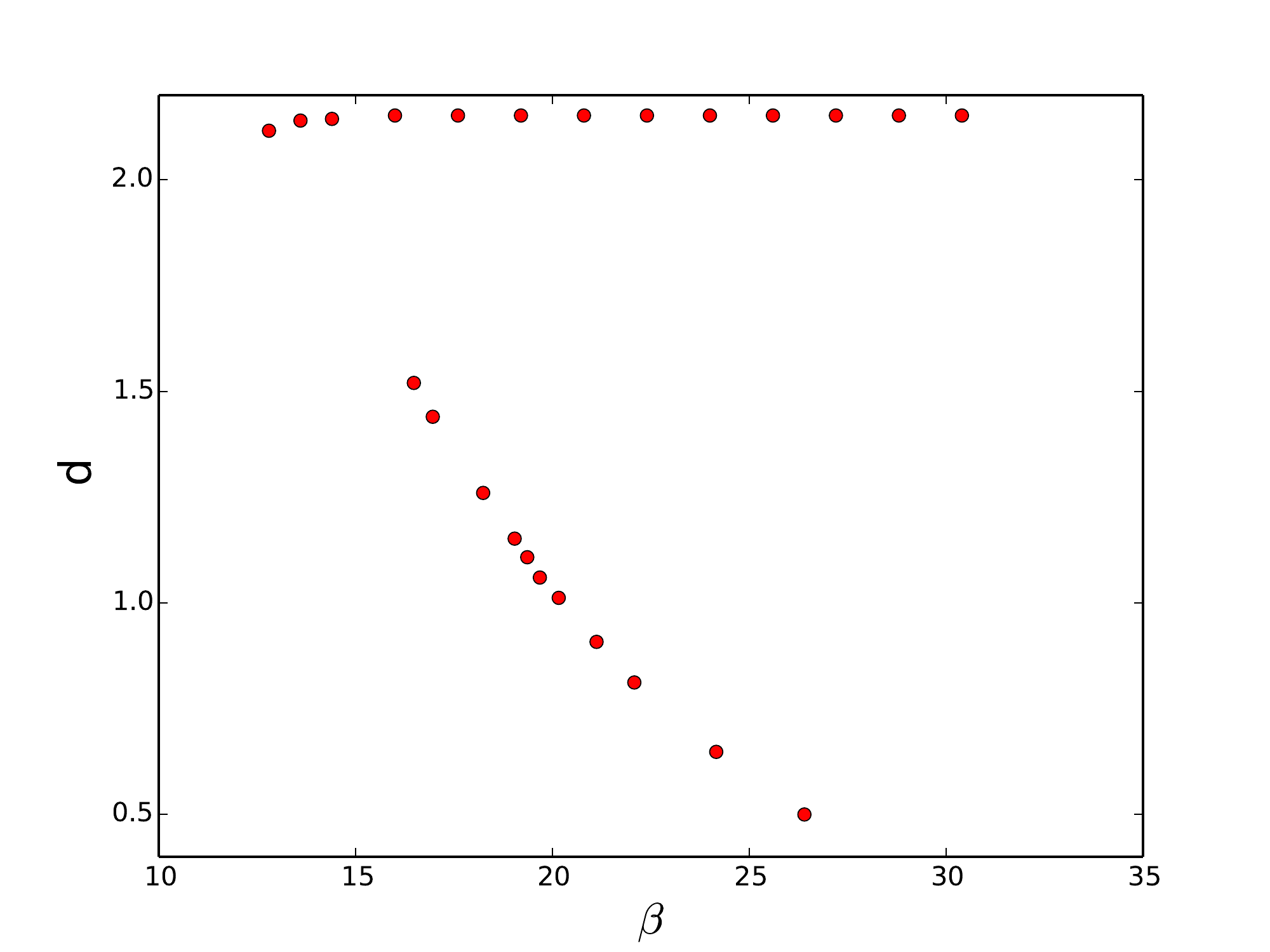}
\caption{\label{f11} Dipole moment of states with broken symmetry and of degenerate states. Measured in units of $M_{\rm cluster}a_0$, the maximum possible value is $2.25$.}
\end{figure}

\begin{figure}[h!]
\centering
\includegraphics[width=.51\textwidth]{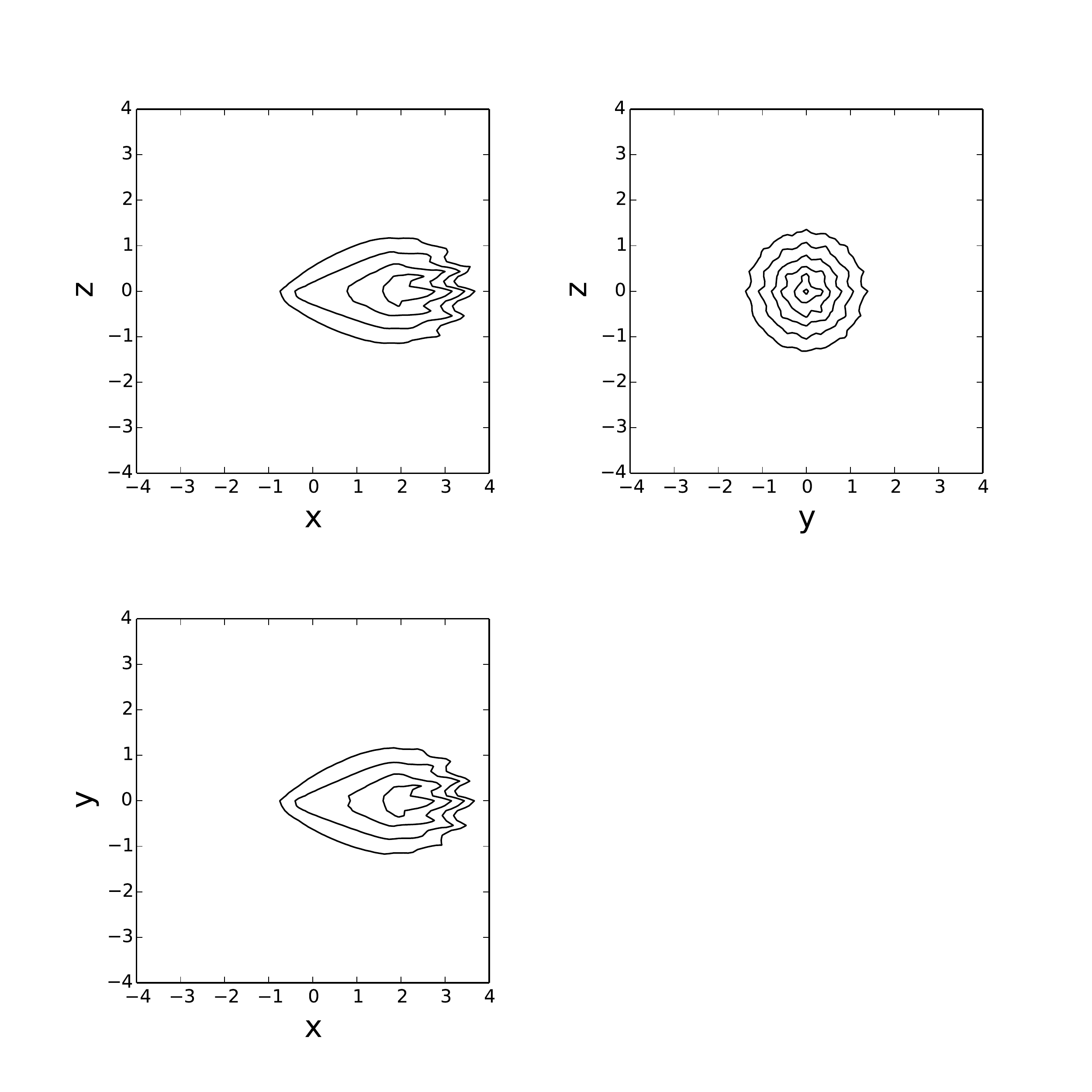}
\caption{\label{f16} Projected surface density of a non-rotating cluster in a lopsided equilibrium, for $\beta=16.5$, $\gamma=0$, $d=1.51$, and $|U|=0.387$.}
\end{figure}

\begin{figure}[h!]
\centering
\includegraphics[width=.51\textwidth]{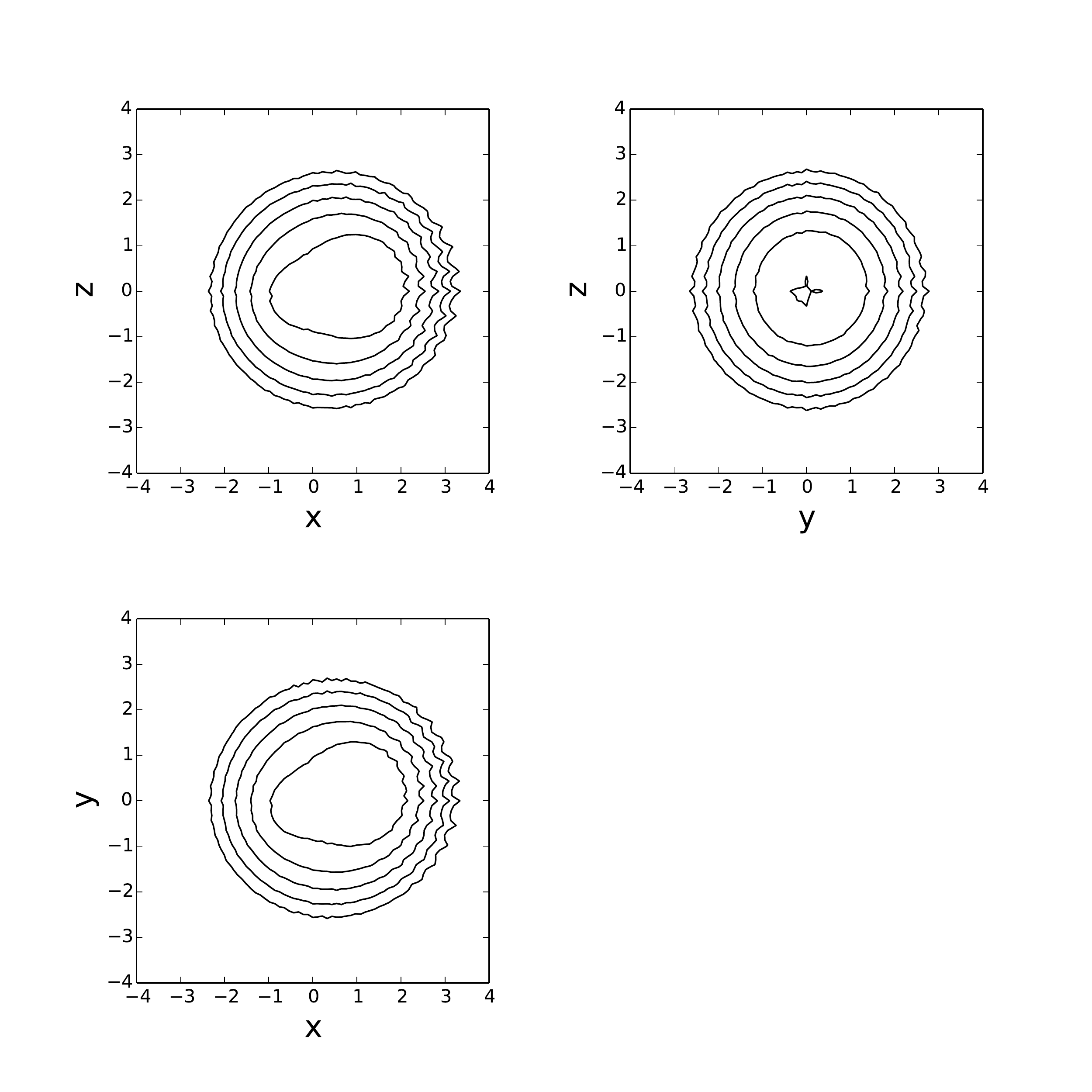}
\caption{\label{f26} Projected surface density of another non-rotating cluster in lopsided equilibrium, for $\beta=25.6$, $\gamma=0$, $d=0.552$, $|U|=0.276$.}
\end{figure}

\begin{figure}[h!]
\centering
\includegraphics[width=.51\textwidth]{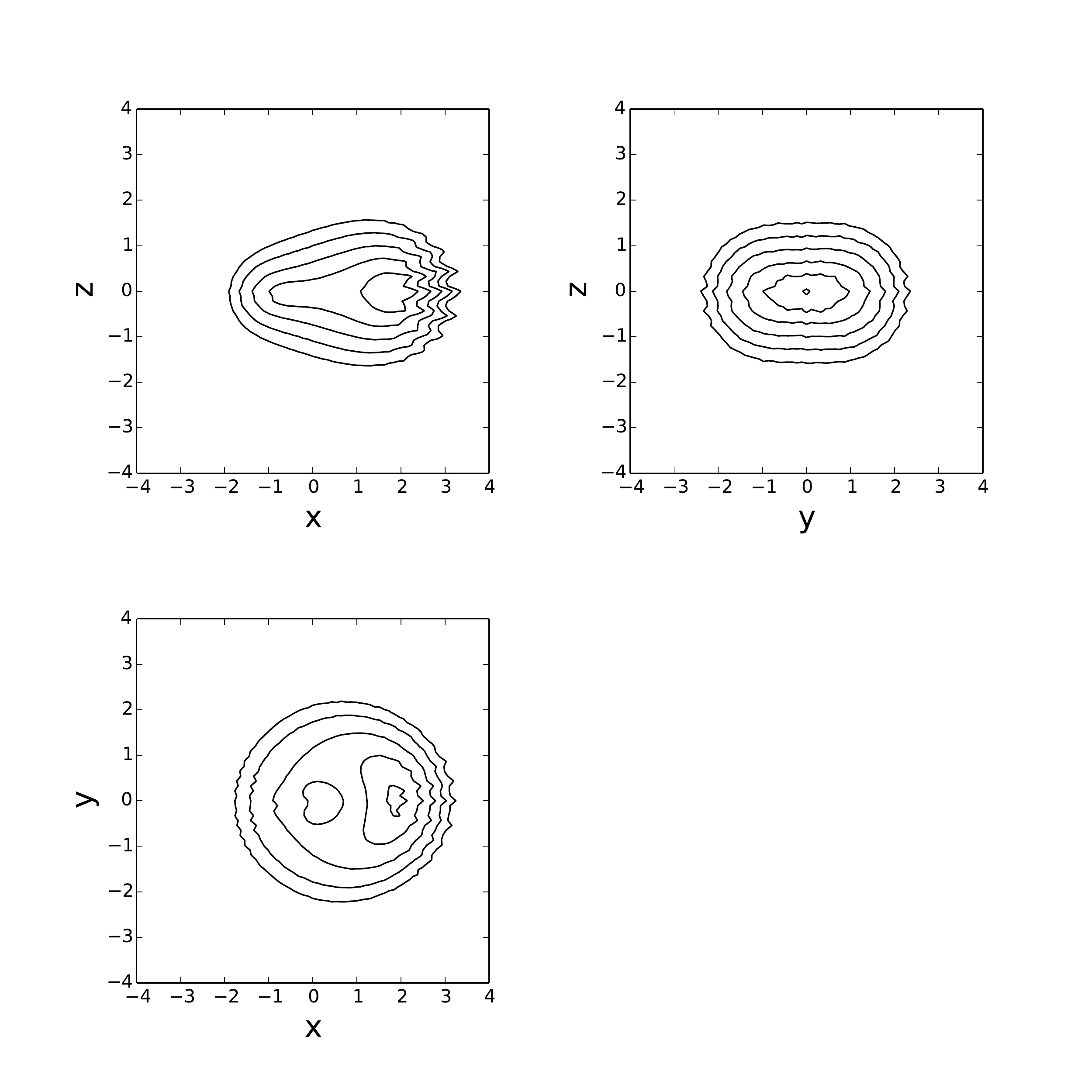}
\caption{\label{f23} Projected surface density of a rotating stellar cluster in lopsided equilibrium, with $\beta=23.0$, $\gamma=2.30$, $d=0.944$, $|U|=0.321$}
\end{figure}

\begin{figure}[h!]
\centering
\includegraphics[width=.51\textwidth]{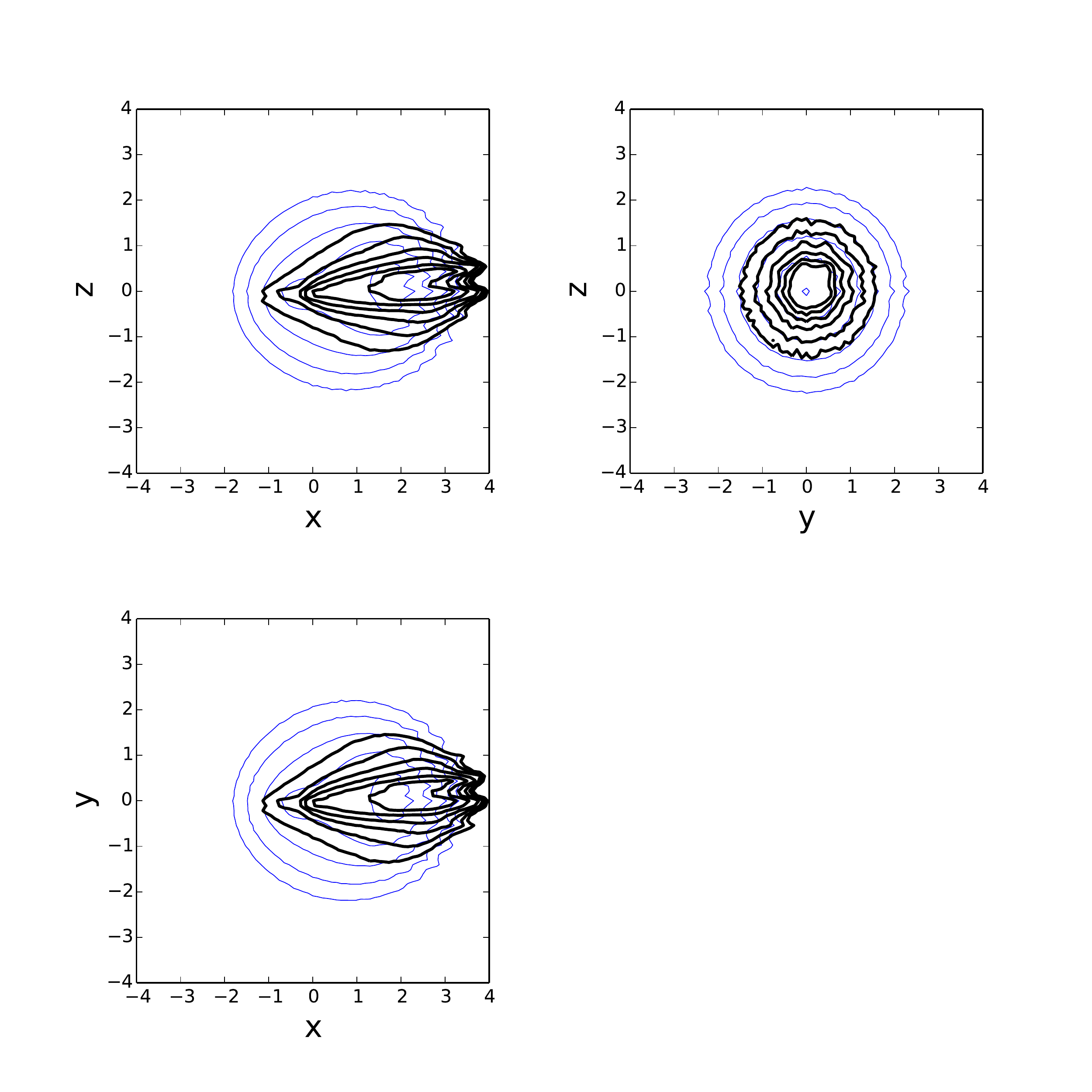}
\caption{\label{f21} Nondegenerate broken-symmetry state with stellar mass black holes, with masses of $10m_0$. The black holes are on more eccentric orbits than the background stars. Thin blue -- background stars. Thick black -- stellar mass black holes. Projected surface density is plotted for $\beta=20.6$, $\gamma=0$, $d=0.972$, $|U|=0.298$. }
\end{figure}

\begin{figure}[h!]
\centering
\includegraphics[width=.51\textwidth]{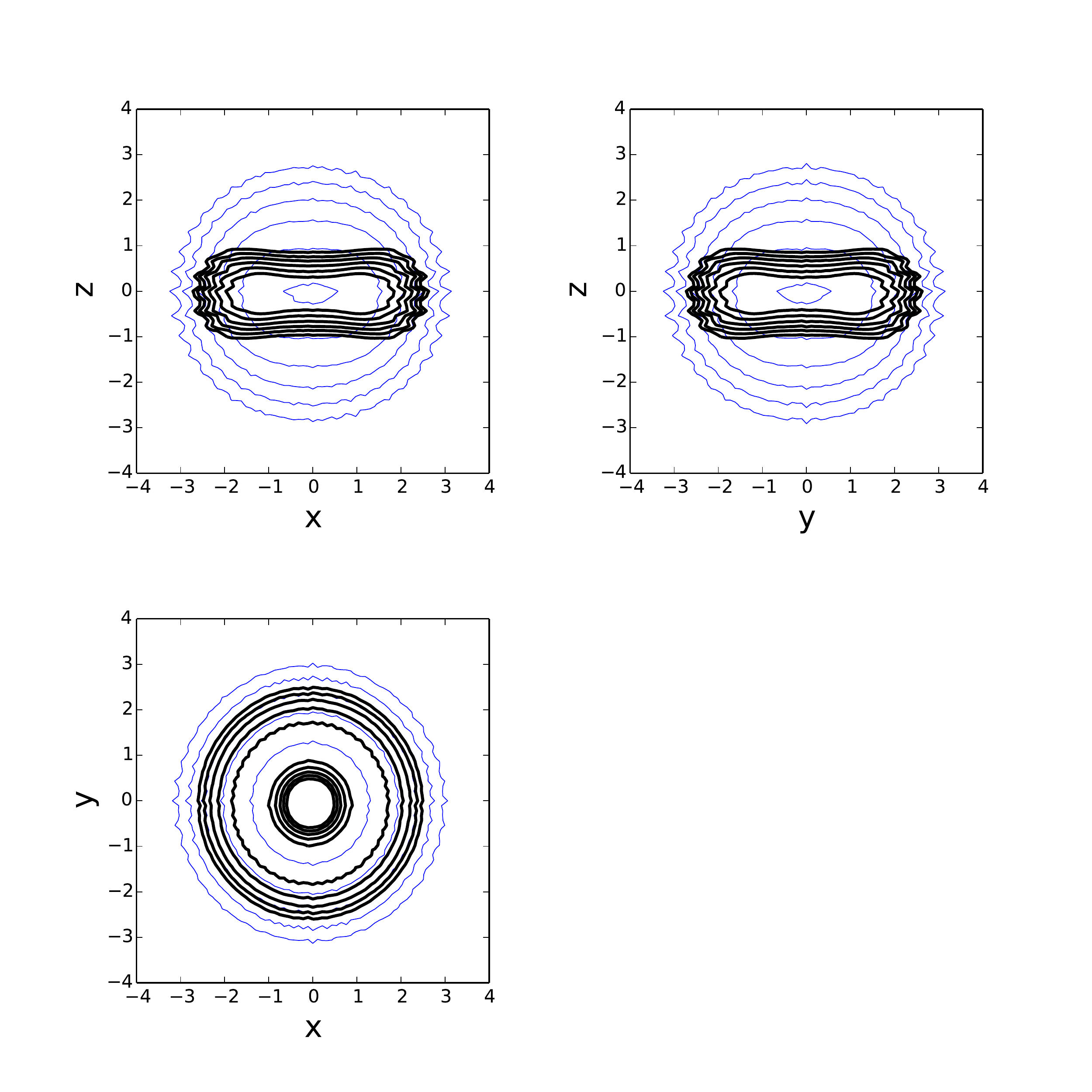}
\caption{\label{f00}  Axially symmetrical rotating state with stellar mass black holes. The black holes are concentrated  near the equator, on more circular orbits.  Projected surface density is plotted for $\beta=0$, $\gamma=2.30$, $d=0.092$, $|U|=0.252$. Projected surface density of stellar-mass black hole sub-cluster inside a rotating axisymmetric cluster is plotted in black.}
\end{figure}


\section{The Numerical Results}\label{seq:numres}

The numerical results described here: (1) demonstrate the  existence of both rotating and non-rotating equilibria with symmetry breaking, (2) give supporting evidence for the existence of the high-$\beta$ singularity and associated degenerate states, (3) show an important effect, that massive objects such as stellar mass black holes occupy special orbits in the clusters.

The examples shown in this section use mostly single-mass $m_0$ star clusters with the uniform distribution of semimajor axes $a_0<a<2a_0$. When stellar-mass black holes are introduced, their mass is $10m_0$. The inverse temperature $\beta$ is measured in units of $\left(GM_{\rm cluster}m_0/a_0\right)^{-1}$, $\gamma$ is measured in units of $\left(\sqrt{GMa_0}m_0 \right)^{-1}$, the binding energy $|U|$ is measured in units of $(GM_{\rm cluster}^2/a_0)$, and the dipole moment $d$ is measured in units of $M_{\rm cluster}a_0$.

\subsection{Non-rotating clusters}
Figure~(\ref{f1}) shows the equilibria for non-rotating clusters we were able to find in the $\beta-|U|$ plane. There are several notable features on this plot:

The lower branch represents spherically symmetric clusters, with the left end (low $\beta$ and low binding energy) featuring very eccentric orbits and the right end (high $\beta$ and higher binding energy) featuring orbits close to circular. A projected density profile of two examples of the spherical equilibria is shown in Figures~(\ref{f24}) and (\ref{f35}). The energy of spherical clusters with purely circular and purely radial orbits can be computed analytically, and we checked that these values are in good agreement with the asymptotic values on our plot. Our procedure for exploring this branch was as follows. 
We start  with a small inverse temperature $\beta\ll -1$, choose a spherically symmetrical initial potential $\phi$, say $\phi=0$, and the program soon saturates in a spherically symmetrical thermodynamic equilibrium with very eccentric orbits.  We then increase $\beta$ gradually and use the  potential computed in the previous step as an initial potential for our iterative procedure described in the previous section. We have also written an independent code that computes equilibria with enforced 
spherical symmetry (and thus has  very high resolution), and we checked that the energy values agree between the $2$ codes. 

When we reach the maximum value of $\beta\simeq 40$,  the algorithm fails to
find a spherically symmetric equilibrium and instead the solution jumps to the upper branch  that we mark as ``degenerate''. As shown in Fig.~(\ref{f13}), the orbits in this state are nearly-radial and strongly aligned, with needle-like projected surface density. We believe that this state represents the high-$\beta$ singularity identified in 
Section 3; obviously with our numerical resolution we do not obtain $|U|=\infty$. The dipole moment  of the degenerate  branch is displayed in Fig.~(\ref{f11}) to be close to the  theoretical maximum value of $9/4$, obtained for perfectly aligned degenerate ellipses with axes uniformly distributed between $1$ and $2$.
We follow the degenerate branch to the left by decreasing $\beta$ in steps and using the potential from the previous step as an initial potential for the iterative procedure. Once we reach the left-most point,  the solution jumps back down to the spherical branch. We have checked that the actual $U$-values for the degenerate branch are very strongly resolution-dependent, as they should be.

Of particular interest is the branch that bifurcates upwards at $\beta\simeq 30$ and $|U|\simeq 0.27$ from the spherical branch.
These are the non-degenerate states with broken symmetry. Two examples of such states are shown in Figs.~(\ref{f16}) and (\ref{f26}). These equilibria are difficult to find, since for a fixed $\beta$ and arbitrary initial potential the solution tends to converge onto the upper or lower branch. Instead of fixing $\beta$, we introduced a feedback loop where we changed $\beta$ every iterative step depending on the current value of the dipole moment or  the binding energy $|U|$. The basic idea is that if the dipole moment becomes large we reduce $\beta$, and if it becomes small we increase it. To obtain the results shown in Figure~(\ref{f1}), we used the following prescription found by trial and error: $\beta_i=40-c~d_{i-1}$,
where $i$ is the index labeling the iterations, $d_{i-1}$ is the dipole moment obtained in the previous iteration and $c$ is a constant. Starting with $c\simeq 20$, and initial $\phi$ with $|\nabla\phi|\simeq 1$, we get a convergent solution that satisfies the extra constraint $\beta=40-c~d$. Then by varying $c$ we obtain part of the non-degenerate broken-symmetry branch that is shown in the figure.
We emphasize that the presence of the feedback loop does not change the fact the program finds a solution of the nonlinear Poisson Eqs.~(\ref{eqdi},~\ref{poiss}), because the program does saturate, meaning that the inverse temperature ultimately becomes a constant. 
This procedure allowed us to find equilibria with broken
symmetry with binding energies up to $|U|=0.40$, but the algorithm failed  to converge for higher energies. We know from Section 3 that equilibria with arbitrarily high binding energies must exist, and therefore we conclude that our failure to find such equilibria are due to computational difficulties and does not reflect a matter of principle.

It is important to remember that a cluster we are considering is  represented by a microcanonical ensemble, with conserved binding energy $|U|$. There is a range of values $0.27\lesssim |U|<1-\ln 2\approx 0.31$ (the theoretical maximum bindidng energy of a spherical cluster) where we are finding solutions with $2$ possible values of $\beta$, one spherically symmetric and one with broken spherical symmetry. It is likely that one of these solutions is meta-stable
(like overheated water or over-cooled water vapour), or  unstable. Intuitively it seems likely that since the broken-symmetry state has
higher temperature, it occupies greater volume of phase space. Therefore it is the broken-symmetry state that is stable. This argument is in harmony with results of \cite{2005ApJ...625..143T} who showed that spherically-symmetric clusters with preferentially circular orbits are subject to secular-dynamical instability.

Finally it is interesting to note that the clusters on the broken-symmetry branch have negative heat capacity. This can potentially lead to thermo-gravitational instability if the cluster comes into contact with the heat bath at the same temperature (how this would be implemented in practice is another matter); presumably in this case the cluster would collapse to a degenerate state.

\subsection{Rotation}

Figure (\ref{f23}) shows an example of a lopsided equilibrium of a rotating cluster. The orbits are eccentric and their eccentricities are strongly aligned with each other.  Notably the surface density in the 
equatorial plane shows two enhancements: one near the supermassive black hole due to the clustering of the stars at small radii due to their $a$-distribution, and the other one due to clustering of the apocenters of orbital ellipses. The nuclear
cluster in Andromeda has similar structure which led \cite{1995AJ....110..628T}
to model it as an ``eccentric disc''. The disc consists of old stars stars \citep{2005ApJ...631..280B} and is likely dynamically old, so one may expect it to reach secular-dynamical equilibrium. It would therefore be of interest to fit the data in Andromeda using rotating lopsided equilibria that we are finding; this is a subject for future work.

We can find the lopsided rotating equilibria by starting with the non-rotating lopsided equilibrium with $\gamma=0$ and then slowly switching on the rotation by incrementally increasing $\gamma$. We find that for sufficiently rapid rotation the cluster becomes axially symmetric; this must take place when the angular velocity of the cluster $\Omega=\gamma/\beta$ exceeds the possible angular velocity of precession of elliptical orbits of the cluster. An example of an axisymmetric rotating cluster is shown in 
Fig.~(\ref{f00}).

\subsection{Stellar-mass black holes}

It is of great astrophysical interest to consider the orbits of heavy objects in a black-hole cluster, such as those of stellar mass black holes. In thermodynamics heavy particles  occupy the lowest available potential energy states. This however, is only true for positive temperatures, so we should be careful: for negative temperatures, the opposite is true. Moreover, extending our intuition from $\beta$ to $\vec{\gamma}$, we may expect that stellar mass black holes will maximally align their angular momenta with the latter.

The lopsided equilibria of the previous subsections take place at positive temperature, therefore black holes will tend to adjust their orbits to minimize their potential energies. This means their eccentricity vectors are expected to be strongly aligned with the lopsidedness of the potential, and their density distribution should be more lopsided than that of the lighter stars. This is demonstrated in Fig.~(\ref{f21}). 

To demonstrate the orbital angular momentum alignment, in Fig.~(\ref{f00}) we show the black hole subcluster of a rotating axisymmetric cluster. While the cluster is only mildly flattened by the rotation, the black hole orbits condense into a disc. This interesting behavior of black holes in rotating nuclear clusters was predicted by \cite{PhysRevLett.121.101101} using a different technique, and is discussed in some detail in the next section.
\vskip .2in

\section{Orbits of stellar-mass black holes: analytical treatment}

At the end of the previous section we saw that stellar-mass black holes are very sensitive ``thermometers" of the clusters; their orbital eccentricity and rotation are strongly amplified compared to the lighter members of the cluster, for those clusters that have broken symmetry or are rotating.
It is possible to gain an analytical handle on this property of the stellar mass black holes, by considering several limiting cases.
\subsection{Spherical clusters}
We explore the case when the background cluster is spherically symmetric and contains a large number of stars. In this case, the overall potential per unit mass has a dominant spherically symmetric smooth component $\phi(r)$, where $r$ is the distance to the supermassive black hole at the center. The fluctuating non-spherical part of the potential leads to the exchange of energy between different orbits, and drives the system to thermodynamic equilibrium. However, only the smooth component is contributing when  evaluating  the Boltzmann weights.

 We will consider general spherical clusters and also, for concreteness, the special case of self-similar (power-law density) clusters. To understand the behavior of heavy stars, we need to analyze the properties of the mean potential energy of an orbit in Eq~(\ref{orbitenergy}). It is given by
\begin{equation}
    u(m,a,l)={m\over 2\pi a}\int\limits_0^{2\pi}\phi[R(l,\xi)]R(l,\xi)d\xi,
    \label{average}
\end{equation}
where 
\begin{equation}
    R=a\left[1-\sqrt{1-l^2}\cos(\xi)\right]
\end{equation}
is the radius.
Here $l=\sqrt{1-e^2}=j\left(GMm^2a\right)^{-1/2}$ is the dimensionless angular momentum of the orbit.
Consider a self-similar spherical clusters with the power-law density distribution,
\begin{equation}
    \rho(r)=C r^{-\delta},
\end{equation}
where $C$ is a constant and $\delta$ is typically between $1.25$ and $1.75$. The potential is then
\begin{equation}
    \phi(r)={4\pi G C\over (3-\delta)(2-\delta)}r^{2-\delta}.
    \label{potentialgamma}
\end{equation}
The  orbit-averaged potential energy is given by
\begin{equation}
    u(m,a,l)={4\pi m G C a^{2-\delta}\over (3-\delta)(2-\delta)}l^{3-\delta}P_{3-\delta}(1/l),
    \label{legendre}
\end{equation}
where $P_\mu$ is the Legendre function. Since the order of the Legendre function is typically non-integer, the expression above is neither intuitive nor very useful. We found it more convenient
to expand it in powers of $l^2$. For example, for the Peebles-Young cusp with $\delta=1.5$, an excellent approximation is 
\begin{eqnarray}
    u(m,a,l)&=&{16\pi m G C a^{0.5}\over 3}\times\\
    & &\left[{8\sqrt{2}\over 3\pi}-{1\over \pi\sqrt{2}}l^2+0.0247~l^4\right].\nonumber
    \label{gamma1.5}
\end{eqnarray}
The first two terms on the right-hand side are obtained analytically  from the Taylor series, while the third term was chosen to match the exact expression at the maximum value of $l=1$. The overall approximation has fractional accuracy better than $3\times 10^{-3}$ for all $l$.

The quadratic dependence on $l$ for small values of $l$ holds for general spherically-symmetric clusters and follows directly from Eq.~(\ref{average}).
One can show  that for $l\ll 1$, 
\begin{equation}
    u(m, a,l)=u_0(m,a)-{1\over 2}\left({GMm\over a}\right)~ \alpha l^2.
    \label{lowl}
\end{equation}
Here the dimensionless coefficient $\alpha$ is positive for $\phi(R)$ created by a stellar cluster, and is given by the following expression:
\begin{equation}
    \alpha={a\over GM}{1\over \pi}\int\limits_0^\pi {d\phi[R(\xi)]\over dR}R(\xi) d\xi,
    \label{lowl1}
\end{equation}
where $R(\eta)=a[1-\cos(\xi)]$ follows that of the radial orbit with semimajor axis $a$. For the potential given by Eq.~(\ref{potentialgamma}), 
\begin{equation}
    \alpha={4\pi C a^{3-\delta}\over M(3-\delta)}{2^{2-\delta}\Gamma(2.5-\delta)\over \sqrt{\pi}\Gamma(3-\delta)}.
    \label{alpha1}
    \end{equation}
It is instructive to write the above equation in terms of $M_{\rm cluster}(<a)$, the mass in stars at radii less than $a$:
\begin{equation}
    \alpha=q(\delta){M_{\rm cluster}(<a)\over M},\label{alpha2}
\end{equation}
where
\begin{equation}
    q(\delta)={2^{2-\delta}\Gamma(2.5-\delta)\over \sqrt{\pi}\Gamma(3-\delta)} .
    \label{alpha3}
\end{equation}
In the range of interest the numerical pre-factor $q$ is not a sensitive function of $\delta$, and it approximately equals $0.9$ for $\delta=1.5$.



It is now straightforward to write down the probability distribution function for an orbit with a semimajor axis $a$ and mass $m$ to have a square eccentricity   $e^2={1-l^2}$:

\begin{equation}
    \mathscr{P}_{a,m}\left({e^2}\right)={1\over N_0}\exp\left[-{\beta u(m,a,l)}\right],
    \label{probability}
\end{equation}
where 
\begin{equation}
    N_0=\int\limits_0^1 d(l^2) \exp\left[-{\beta  u(m,a,l)}\right]
    \label{normalization}
\end{equation}
is the normalization factor.

For fixed $a$ and $m$, the variation of $u(m,a,l)$   is approximately given by
\begin{equation}
    u(m,a,0)-u(m,a,1)\sim \alpha{GMm\over 2a},
    \label{variation}
\end{equation}
where $\alpha$ is given by Eq.~(\ref{lowl1}); for the power-law cluster, $\alpha$ is given by Eq.~(\ref{alpha2}). Therefore the character of the $l$-distribution (and therefore the
character of the eccentricity distribution) is determined by a dimensionless parameter
\begin{equation}
    \bar{\beta}(m,a)={\beta GMm\alpha\over 2a}.
    \label{betabar}
\end{equation}
There are 3 limiting cases:

Case 1: $\left|\bar{\beta}\right|\ll 1$. In this high-temperature limit, the distribution is uniform in the $l^2$, and 
\begin{eqnarray}
    \mathscr{P}(l)&=&2l\nonumber\\
    \mathscr{P}(e)&=&2e.
    \label{thermal}
\end{eqnarray}
For historical reason, this is called the ``thermal" distribution of eccentricities and angular momenta. In fact, a more accurate name is the maximum-entropy distribution. While it is {\it assumed} to hold for relaxed clusters in much of the literature on resonant relaxation, we emphasize that it is really the high-temperature subset of possible thermal equilibria. For $\bar{\beta}<0$, i.e.~for negative temperature, the values of
$l$ will on average be lower than those of the distribution in Eq.~(\ref{thermal}), and thus the orbits will be more eccentric. Conversely, for $\bar{\beta}>0$, i.e. for positive temperature, the orbits will on average be less eccentric than those in Eq.~(\ref{thermal}). The other two limiting cases are

Case 2: $\bar{\beta}\ll - 1$. In this low negative temperature limit, the orbits are eccentric and the distribution is exponential in $l^2$, given by
\begin{equation}
\mathscr{P}\left(l^2\right)\simeq{\left|\bar{\beta}\right|}\exp\left[{-\left|\bar{\beta}\right| l^2}\right].
\label{maxwell}
\end{equation}
The associated mean values are
\begin{eqnarray}
    \langle l^2 \rangle&=&\left|\bar{\beta}\right|^{-1}\nonumber\\
    \langle e^2 \rangle&=&1-\left|\bar{\beta}\right|^{-1}.\label{lowT+}
\end{eqnarray}
The analysis in this paper and in ${\rm TT}$ shows that such ``hedgehog'' clusters are stable.

Case 3: $\bar{\beta}\gg 1$. In this low positive temperature limit, the orbits are nearly circular (tangential). The computations in ${\rm TT}$ and this paper suggest that the clusters with preferentially tangential orbits are unstable and develop strongly lopsided structures. 

\paragraph{Power-law cusps.}
Consider as a useful example the power-law cusp, with 
$\alpha$ given by Eq.~(\ref{alpha1}). In that case
\begin{eqnarray}
    \bar{\beta}(a)&=&{\beta G m C a^{2-\delta}}{\sqrt{\pi}2^{3-\delta}\Gamma(2.5-\delta)\over (3-\delta)\Gamma(3-\delta)}\nonumber\\
    &=&{\beta\over |\beta|}{m\over m_0}\left({a\over a_{\rm th}}\right)^{2-\delta},
    \label{barbeta2}
\end{eqnarray}
where $m_0$ is the mass of a typical star in the cluster, and 
\begin{equation}
    a_{\rm th}= \left[|\beta|G m_0 C {\sqrt{\pi}2^{3-\delta}\Gamma(2.5-\delta)\over (3-\delta)\Gamma(3-\delta)}\right]^{1\over \delta-2}.
\end{equation}
is the semimajor axis at which $|\bar{\beta}|=1$ for a star of mass $m_0$.
We see immediately that for $a\ll a_{\rm th}$, the orbits of stars with mass $m_0$ are following the maximum-entropy distribution of Eq.~(\ref{thermal}).
For $a>a_{\rm th}$ the orbits are becoming more eccentric  as $a$ increases, if the temperature is negative. For positive temperature, the orbits become more circular as $a$ increases, and the cluster is likely to develop a lopsided configuration beyond
some critical radius, thus breaking the spherical symmetry. 

\paragraph{Heavy objects inside spherical clusters}
Black holes  as well as massive stars likely exist inside  nuclear star clusters, and their masses can be much greater than those of the majority of the cluster members. From Eq.~(\ref{barbeta2}), we see that the dimensionless temperature parameter $\bar{\beta}$ scales linearly with the mass of the object. The heavy objects will have a different eccentricity distribution than the majority of
the stars with the same semimajor axes. In fact, for $\delta=1.5$, a black hole with the mass $10$ times greater than the average stellar mass, will have the same eccentricity distribution as the majority of the stars with the semimajor axes $100$ times greater than the black hole's!

In other words, for a negative-temperature cluster, heavy objects are on more eccentric orbits than their neighbours; this is the effect that was likely seen in numerical experiments of \cite{2012ApJ...754...42M}. Conversely, for positive temperatures (if the cluster is still stably spherical), the heavy objects are on less eccentric orbits than their neighbours. The effect can be quite dramatic, as illustrated in Fig.~{\ref{meaneccentricity} where mean eccentricities are plotted for stars and black holes in a power-law cluster with $\delta=1.5$. We see that if the mean eccenricity  of the stars exceed that of the maximum-entropy distribution, the orbits of black holes and other heavy objects are substantially more eccentric than those of the rest of the stars, on average. Conversely, if the orbits of the background stars are more circular on average than $e=2/3$, then the orbits of  black holes and other heavy objects are substantially more circular than those of the background stars. 

\begin{figure}[t]
\centering
\includegraphics[width=.51\textwidth]{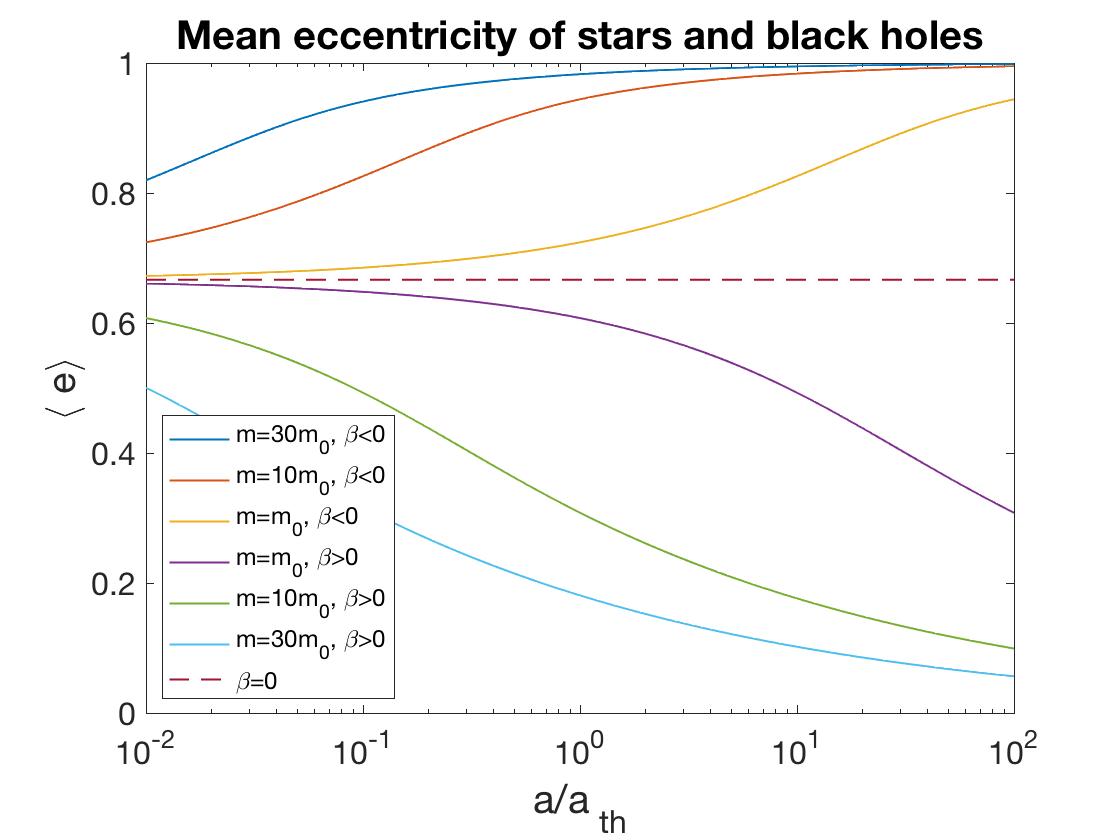}
\caption{Mean eccentricity as a function of radius for a power-law cluster with $\delta=1.5$, plotted for populations with $3$ characteristic masses: $m=m_0$ (the background stars), $m=10m_0$, and $m=30m_0$ (the black holes and massive stars)}
\label{meaneccentricity}
\end{figure}

The eccentricity of the black-hole orbits is a sensitive thermometer for the rest of the stellar distribution. We can see near $a=a_{\rm th}$, where the background distribution deviates very slightly from the maximum-entropy one, the heavy black holes amplify dramatically these deviations. Clearly this will have major consequences on the interaction of heavy objects with the supermassive black hole, since 
these interactions require the heavy object acquiring an extremely eccentric orbit. We will postpone the detailed discussion of such interactions to future work, since they require understanding of not just an equilibrium distribution but also the stochastic evolution of the orbits.

\subsection{Rotation}

Rotation impacts the distribution of stellar orbits in two ways. It introduces a second temperature-like parameter $\vec{\gamma}$ that enters into the Boltzmann weight 
through a factor $\exp\left[\vec{\gamma}\cdot {\bf j}\right]$. This extra factor creates a preference for the angular momenta of the 
stars to be aligned with $\vec{\gamma}$. Rotation also flattens the cluster towards its equatorial plane, via the direction-averaged Boltzmann factor $\cosh\left[\vec{\gamma}\cdot {\bf j}\right]$. Since $j\propto m$, stellar mass black holes' angular momenta are much stronger aligned 
than those of the rest of the stars, and as we saw in Section 5, for realistic parameters they form a disc-like configurations inside rotating clusters.

Without loss of generality we choose the $z$-axis to be aligned with $\vec{\gamma}$. We work with Delaunay action-angle variables for Keplerian orbits with fixed semimajor axes, with actions $j$, $j_z$ and corresponding angles $\zeta$, $\zeta_z$. Here $\zeta_z$ is the angle of the line of nodes\footnote{A common term in celestial mechanics, signifying the angle between the $x$-axis and  the line of intersection between the 
orbital plane and the $x-y$ plane} and $\zeta$ is the argument of the periastron\footnote{The angle between the line of nodes and the  radial line through the periastron of the orbit.}.
The probability distribution function for a star with fixed $a,m$ is given by
\begin{equation}
    \mathscr{P}_{m,a}(j,j_z,\zeta, \zeta_z)={1\over N_2}\exp\left[-\beta u(m,a,j,j_z,\zeta,\zeta_z)+\gamma j_z\right].
\end{equation}
Here as always $u$ is the orbit-averaged energy, $j$ is restricted to vary between $0$ and $j_c=m\sqrt{GMa}$,  $j_z$ is restricted to vary between $-j$ and $j$, and $N_2$ is the normalization.

In general, the potential energy $u$ has to be computed numerically as was done in Section 5. To gain intuition from
an analytical calculation, we consider $3$ limiting cases below:

\paragraph{Case 1: heavy black holes with $m\gg m_0$.} Since both $u$ and ${\bf j}$ scale linearly with $m$, such black holes 
will cluster around the orbit that maximizes the function 
\begin{equation}
    p(j/m,j_z/m,\zeta,\zeta_z)\equiv \left({1/m}\right)\left[{\gamma} {j_z}-\beta u\right].
    \label{p}
\end{equation}
This implies that $\partial u/\partial \zeta=0$ and $\partial u/\partial \zeta_z=0$, so the orbit experiences no torque along
${\bf j}$ or $z$-axis. If the orbit is inclined, the torque 
\begin{equation}
    \vec{\tau}\propto \hat{z}\times{\bf j}
\end{equation}
and the angular momentum vector ${\bf j}$ precesses around the $z$-axis. How quickly  would it precess? The inclined 
orbit implies $\left|j_z\right|<j$, so maximizing $p$ with respect to $j_z$ gives 
\begin{equation}
    {\partial u\over\partial j_z}=\Omega.
\end{equation}
The left-hand side is the rate of precession of the line of nodes in the $x-y$ plane. Maximizing $p$ with respect to $j$ implies that either $j=j_c$ and the orbit is circular, or $\partial p/\partial j=0 $ and $\zeta$ is constant. In either case\footnote{At a first glance it seems logically possible that $p$ could be maximized at $j=j_z=0$. However, recall that
$\partial u/\partial j=0$ at $j=0$, and therefore this cannot be a maximum of $p$ when $\gamma$ is not zero.} the orbit is stationary in the frame of reference rotating with $\Omega$. 

If the orbit is located in the equatorial plane, $j_z=j$ (for simplicity, we can choose the direction of the $z$ axis to fix the $+$ sign). Maximizing $p$ with respect to $j$ implies that ether the orbit is circular, or
\begin{equation}
    {\partial u(m,a,j,\zeta)\over  \partial j}=\Omega.
\end{equation}
\vskip .1in
Here $\zeta$ is the argument of the periastron in the equatorial plane (relative to e.g., $x$-axis). The orbit is either circular or it precesses with angular velocity $\Omega$ in the equatorial plane. Therefore we proved generally that {\bf heavy black holes are attracted to orbits that are stationary in the frame of reference rotating with angular frequency $\Omega$}.

\paragraph{Case 2: infinite-temperature cluster ($\beta=0$).} The mathematics becomes fully analytical: the angles $\zeta$, $\zeta_z$ drop out and we get
\begin{equation}
    \mathscr{P}_{a,m}(j,j_z)={\gamma^2 e^{\gamma j_z}\over 2\left[\cosh\left(\gamma j_c\right)-1\right]}
    \label{inftemp}
\end{equation}
for $0\le j\le j_c$ and $-j\le j_z\le j$, and $0$ otherwise. The inclination angle of the orbit $0\le\theta\le \pi$ is given by $c\equiv \cos\theta=j_z/j$, and 
\begin{eqnarray}
    \mathscr{P}_{a,m}(c)&=&\int\limits_0^{j_c}dj \int\limits_{-j}^j dj_z\mathscr{P}_{a,m}(j,j_z)\delta\left({j_z/ j}-c\right)\\
    &=& {1+e^{\bar{\gamma}c}\left(\bar{\gamma}c-1\right)\over 2\left(\cosh\bar{\gamma}-1\right)c^2},\nonumber
\end{eqnarray}
where $\bar{\gamma}=\gamma j_c$ fully determines the distribution of inclinations. For slow rotation $\bar{\gamma}\ll 1$ and $\mathscr{P}(c)=1/2$, which corresponds to isotropically distributed orbits. For rapid rotation $\bar{\gamma}\gg 1$ and the values of $c$ are concentrated near $1$, with the probability density approximately given by
\begin{equation}
    \mathscr{P}_{a,m}(c)\simeq \bar{\gamma}e^{-\bar{\gamma}(1-c)}.
\end{equation}
In this limit the inclination angles are concentrated near zero and their probability density is given by
\begin{equation}
    \mathscr{P}_{a,m}(\theta)\simeq \bar{\gamma}\theta~e^{-{1\over 2}\bar{\gamma}\theta^2},
\end{equation}
and the mean value of the inclination is given by
\begin{equation}
    \langle\theta\rangle\simeq\sqrt{\pi\over 2\bar{\gamma}}.
    \label{inclination1}
\end{equation}
Note that the mean inclination angle is weakly decreasing with the semimajor axis, $\langle\theta\rangle \propto a^{-1/4}$ and 
more sensitively decreasing with the black-hole mass, $\langle\theta\rangle\propto  m^{-1/2}$. Finally, we note that rotation makes the orbits on average more circular, with
\begin{equation}
    \mathscr{P}_{a,m}(l)={\bar{\gamma} \sinh\left[\bar{\gamma}l\right]\over \cosh\bar{\gamma}-1}.
\end{equation}
For $\bar{\gamma}\gg 1$, the eccentricity values cluster around zero, with the
probability distribution function
\begin{equation}
    \mathscr{P}_{a,m}(e)\simeq \bar{\gamma} e\exp\left[-{1\over 2}\bar{\gamma}e^2\right],
\end{equation}
and the mean value of eccentricity the same as that of the inclination:
\begin{equation}
    \langle e\rangle\simeq\sqrt{\pi\over 2\bar{\gamma}}.
    \label{eccentricity1}
\end{equation}

\paragraph{Case 3: nearly spherical cluster with rotation.} We saw in the previous paragraph that rotation makes the orbital distribution more circular on average. This effect was demonstrated for $\beta=0$ and is expected to be suppressed or enhanced for $\beta<0$ and $\beta>0$ respectively. To study this we assume that both $\beta$ and $\gamma$ are non-zero but that the potential is spherically symmetric and is given by Eq.~(\ref{potentialgamma}). We must keep in mind that this approximation is not self-consistent for rapidly rotating clusters with high $\gamma$, but it does give us a qualitative picture of the effect of the two temperatures on the distribution of black hole orbits. Furthermore, we specify the density profile to $\delta=1.5$, with the orbit-averaged energy $u(m,a,l)$ given by Eq.~(\ref{gamma1.5}).

\begin{figure}
    \centering
    \includegraphics[width=.51\textwidth]{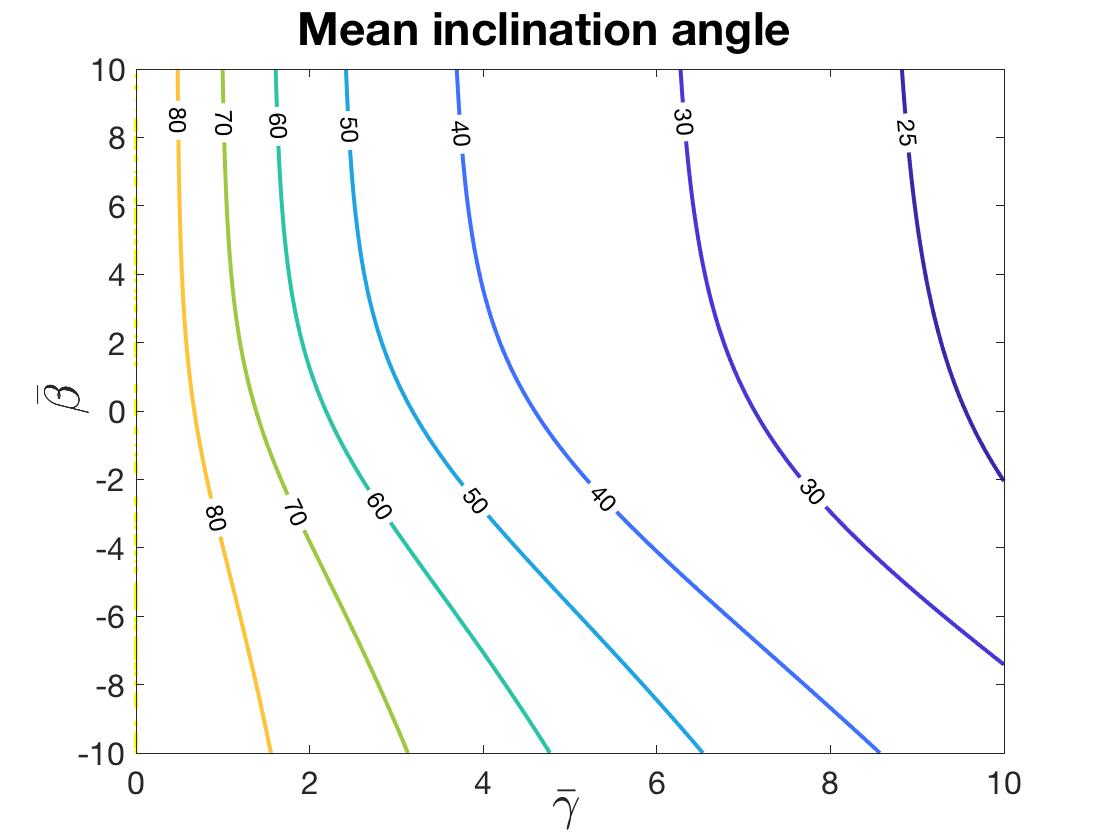}
    \caption{Mean inclination of orbits as a function of $\bar{\beta}$ and $\bar{\gamma}$ for a rotating cluster with $\delta=1.5$. Both parameters scale linearly with mass, so stellar mass black holes tend to have high $\bar{\gamma}$ and their orbits strongly align with the rotation of the cluster.}
    \label{inclination1.5}
\end{figure}

\begin{figure}
    \centering
    \includegraphics[width=0.51\textwidth]{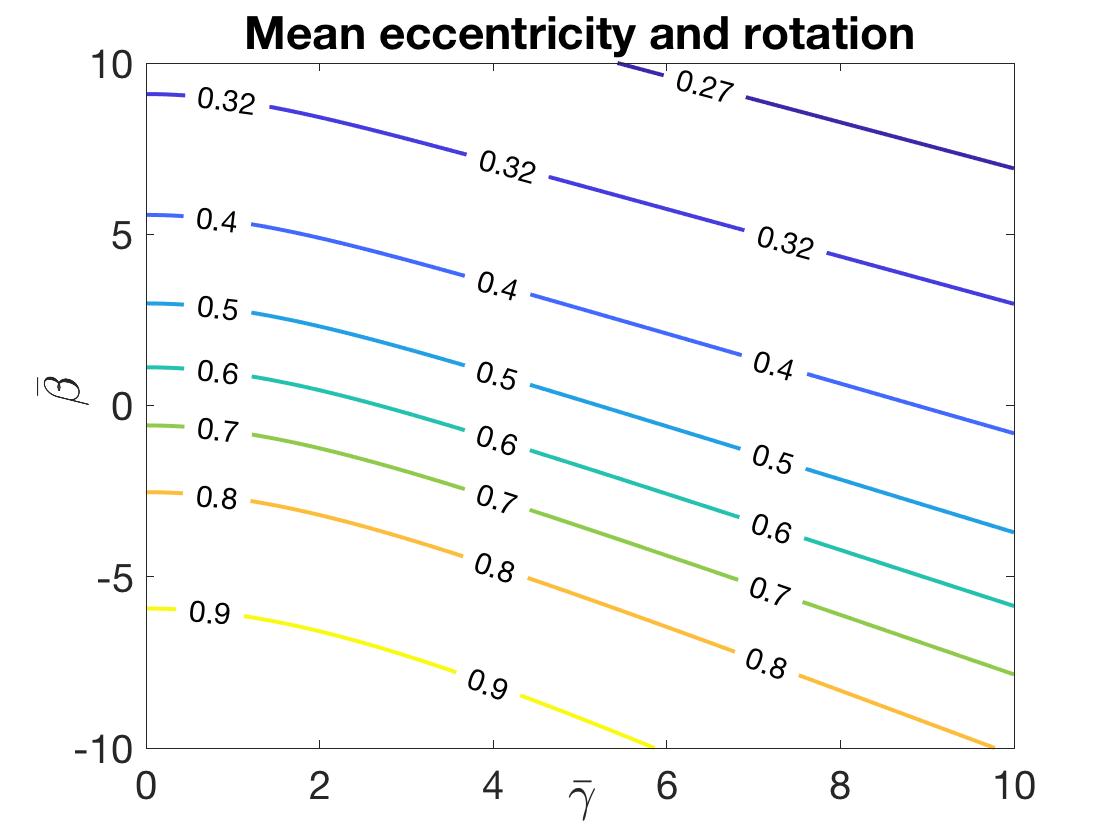}
    \caption{Mean eccentricity of the orbits decreases with rotation, which can have an effect on interaction of stars and stellar mass black holes with the supermassive black hole at the center of the cluster.}
    \label{eccentricity1.5}
\end{figure}

\begin{figure}
    \centering
    \includegraphics[width=0.51\textwidth]{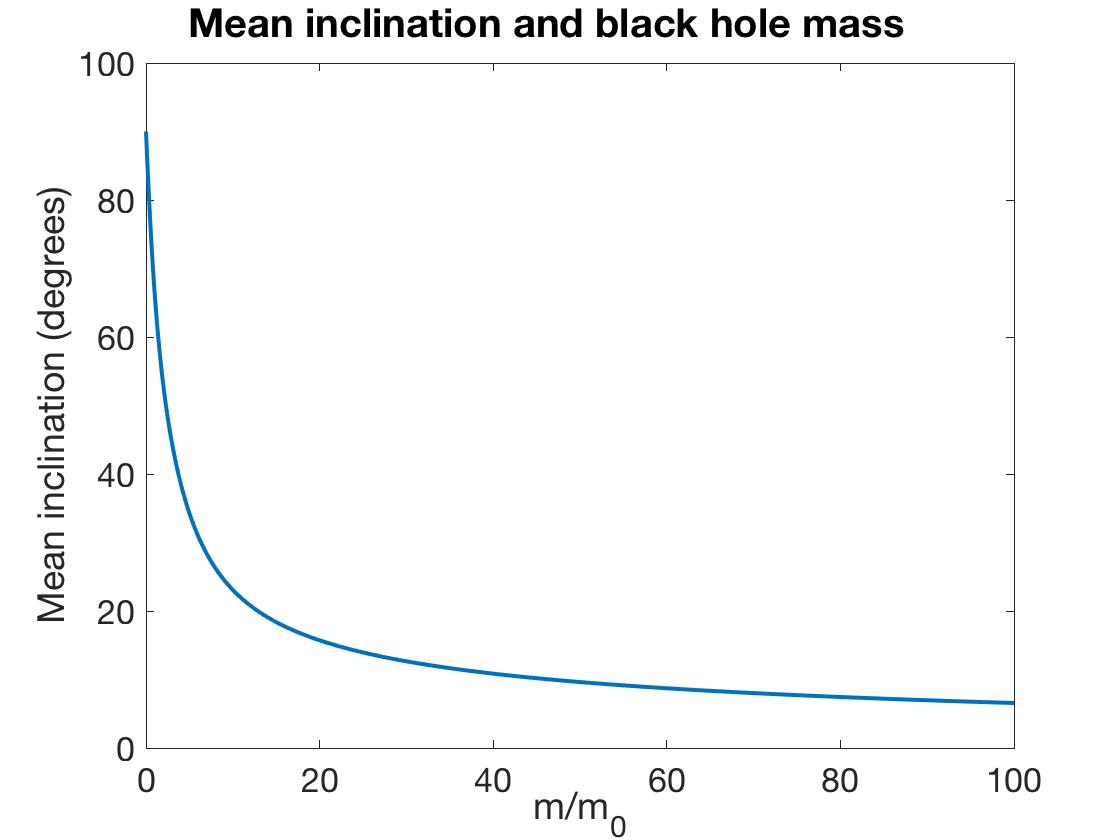}
    \caption{Mean inclination of a black hole as a function of its mass, for $\bar{\gamma}=-\bar{\beta}=2(m/m_0)$.}
    \label{inclinationBH}
\end{figure}

With these assumptions, the probability density distribution for $(l,l_z)=(j/j_c, j_z/j_c)$ becomes
\begin{equation}
    \mathscr{P}_{a,m}(l,l_z)={1\over N_1}\exp\left[\bar{\beta}\left(l^2-0.11~l^4\right)+\bar{\gamma}l_z\right].
    \label{rotprob}
\end{equation}
where $N_1$ is the normalization factor and $\bar{\beta}$ is given by Eq.~(\ref{barbeta2}).
It is worth emphasizing that for a given power-law exponent $\delta$ of the cluster's density profile, the probability distribution function with respect to $l,l_z$ is completely specified by the dimensionless temperature and rotation parameters, $\bar{\beta}$ given by Eq.~(\ref{betabar}), and  $\bar{\gamma}$. The probability distribution above peaks for
aligned orbits with $l_z=l$, which are circular ($l=1$) if $\bar{\beta}\ge -0.64\bar{\gamma}$ and eccentric and precessing with the cluster's angular velocity 
\begin{equation}
    \Omega={\alpha\bar{\gamma}\over 2\bar{\beta}}\sqrt{GM\over a^3}
\end{equation}
if $\bar{\beta}< -0.64\bar{\gamma}$.
In Figures \ref{inclination1.5} and \ref{eccentricity1.5} we show the mean inclination and mean eccentricity of the orbits as a function $\bar{\gamma}$ and $\bar{\beta}$, computed for a cluster with $\delta=1.5$. We can see that the rotational vector $\vec{\gamma}$ biases the orbital angular momenta to be co-aligned with it, and in the high-$\bar{\beta}$ case, the orbits are particularly susceptible to this
co-alignment. Since $\bar{\gamma}$ scales with the mass of the star, {\it the orbits of heavy stars and black holes will align their angular momenta with $\vec{\gamma}$ even for modest rotations of the background clusters}. We believe this argument is consistent with the ``black-hole discs" seen in recent Monte-Carlo simulations with circular orbital annuli by \cite{PhysRevLett.121.101101}
Figure \ref{inclinationBH} illustrates the degree of alignment of  black hole orbits with the the cluster rotation vector, as a function of the black hole mass. As the latter is increased, the orbits get locked into the equatorial plane, as expected.

\subsection{Planets, comets, and other light particles}.

Of some astrophysical interest is the dynamics of very light particles (as compared to the stars) that might be present in galactic nuclei. \cite{2012MNRAS.419.1238N} argue that supermassive black holes are surrounded by swarms of comets and asteroids. \cite{1999PhRvL..83.1719G} show that the growth of a supermassive black hole naturally leads to dark matter spike in its vicinity. In both cases the total  mass of the light particles is subdominant to that of the stars surrounding the black hole, and their gravitational dynamics is determined by that of the stars. Since $\bar{\beta}$ and $\bar{\gamma}$ scale linearly with the mass, $\bar{\beta}=\bar{\gamma}=0$ is a very good approximation. Therefore the light particles are expected to form a spherical sub-cluster with no observable rotation, regardless of how rapidly the stellar cluster rotates and how asymmetric it is. They follow the maximum-entropy distribution in eccentricities, $\mathscr{P(}e)=2e$. This remarkable simplicity should be of use for studies exploring observational signatures of light objects in galactic 
nuclei.


\section{Conclusions}

Touma--Tremaine thermodynamics is a powerful tool for describing the secular-dynamical equilibria of stellar clusters near supermassive black holes. In this paper we give a general analytical and numerical treatment of thermal equilibria, both for non-rotaing and rotating clusters. We show that the existence of lopsided equilibria is robust and argue that the eccentric nuclear disc of Andromeda is likely an example of thermal equilibrium in a rotating precessing cluster. 

We argue that heavy stellar-mass black holes are sensitive ``thermometers'' of the clusters and are attracted to a special set of orbits. For 
spherical non-rotating clusters they are either much more or much less eccentric than the lighter stars, depending on the sign of the temperature. In rotating clusters, they tend to form disc-like structures, as was previously argued by \cite{PhysRevLett.121.101101} by carrying out Monte-Carlo simulations with circular orbital annuli.
Consistent with this, preferentially low inclinations for heavy stars were observed by \cite{2020ApJ...890..175F} in their simulations of an eccentric disc. Additionally, we show that black-hole orbits tend to be stationary in a frame of reference rotating with the cluster's angular velocity. In lopsided clusters, their eccentricity vectors tend to be lined up with the direction of asymmetry of the cluster, more so than those of the lighter stars.  Importantly, cluster rotation tends to deplete strongly eccentric orbits and may reduce 
the amount of stars and especially black holes interacting with the supermassive black hole. This could have a profound impact on the tidal disruption events and on gravitational-wave-driven inspirals of stellar-mass black holes in galactic nuclei. These topics will be explored in future work.

On the other end of the mass spectrum, we remark that comets, asteroids, and dark matter particles form a spherically symmetric non-rotating sub-cluster inside a generally rotating and possibly lopsided black hole cluster. This non-intuitive statement is an immediate consequence of Touma--Tremaine thermodynamics, and should inform studies of observational signatures of such light objects in galactic nuclei.

We thank Scott Tremaine for numerous insightful discussions on stellar dynamics in galactic nuclei, and Jihad Touma for useful feedback on the draft of this paper.


\bibliography{ms4}

\clearpage

\end{document}